\documentclass[lettersize,journal]{IEEEtran}
\usepackage{amsmath,amsfonts}
\usepackage{algorithmic}
\usepackage[algo2e,ruled,linesnumbered]{algorithm2e}

\SetCommentSty{mycommfont}
\usepackage{algorithm}
\usepackage{array}
\usepackage{amssymb}
\usepackage[caption=false,font=footnotesize]{subfig}
\usepackage{textcomp}
\usepackage{stfloats}
\usepackage{url}
\usepackage{verbatim}
\usepackage{graphicx}
\usepackage{hyperref}
\usepackage{multirow}
\usepackage{booktabs} 
\usepackage{cite}
\usepackage{xcolor}
\usepackage{pifont} %
\newcommand{\cmark}{\ding{51}}%
\newcommand{\xmark}{\ding{55}}%

\newcommand{\modify}[1]{{\color{black}#1}}

\begin{document}

\title{USED: Universal Speaker Extraction and Diarization}

\author{Junyi Ao, Mehmet Sinan Y{\i}ld{\i}r{\i}m, Ruijie Tao, Meng Ge, Shuai Wang, \textit{Member, IEEE}, Yanmin Qian, \textit{Senior Member, IEEE}, Haizhou Li, \textit{Fellow, IEEE}

\thanks{This work is supported by China NSFC projects under Grants No. 62401377 and 62271432; Internal Project of Shenzhen Research Institute of Big Data under Grants No. T00120220002 and No.J00220230014; Shenzhen Science and Technology Program ZDSYS20230626091302006; Shenzhen Science and Technology Research Fund (Fundamental Research Key Project Grant No. JCYJ20220818103001002). This work is also supported in part by Huawei Noah’s Ark Lab. (\itshape Corresponding author: Meng Ge).}
\thanks{Junyi Ao, and Haizhou Li are with Shenzhen Research Institute of Big Data, School of Data Science, The Chinese University of Hong Kong, Shenzhen 518172, China (e-mail: junyiao1@link.cuhk.edu.cn; haizhouli@cuhk.edu.cn).}
\thanks{Mehmet Sinan Y{\i}ld{\i}r{\i}m, and Ruijie Tao are with the Department of Electrical and Computer Engineering, National University of Singapore, Singapore 119077 (e-mail: sinan@u.nus.edu; ruijie@nus.edu.sg).}
\thanks{Meng Ge is with Saw Swee Hock School of Public Health, National University of Singapore, Singapore 117549 (e-mail: gemeng@nus.edu.sg)}
\thanks{Shuai Wang is with Shenzhen Research Institute of Big Data, Shenzhen 518172, China (e-mail:wangshuai@cuhk.edu.cn)}
\thanks{Yanmin Qian is with the Auditory Cognition and Computational Acoustics Lab, the Department of Computer Science and Engineering and the MoE Key Laboratory of Artificial Intelligence, AI Institute, Shanghai Jiao Tong University, Shanghai 200240, China (e-mail: yanminqian@sjtu.edu.cn).}
}

\markboth{Journal of \LaTeX\ Class Files,~Vol.~14, No.~8, August~2021}
{Shell \MakeLowercase{\textit{et al.}}: A Sample Article Using IEEEtran.cls for IEEE Journals}

\maketitle

\begin{abstract}
Speaker extraction and diarization are two enabling techniques for real-world speech applications. Speaker extraction aims to extract a target speaker's voice from a speech mixture, while speaker diarization demarcates speech segments by speaker, annotating `who spoke when'. Previous studies have typically treated the two tasks independently. In practical applications, it is more meaningful to have knowledge about `who spoke what and when', which is captured by the two tasks. The two tasks share a similar objective of disentangling speakers. Speaker extraction operates in the frequency domain, whereas diarization is in the temporal domain. It is logical to believe that speaker activities obtained from speaker diarization can benefit speaker extraction, while the extracted speech offers more accurate speaker activity detection than the speech mixture. In this paper, we propose a unified model called Universal Speaker Extraction and Diarization (USED) to address output inconsistency and scenario mismatch issues. It is designed to manage speech mixture\modify{s} with varying overlap ratios and variable number of speakers. We show that the USED model significantly outperforms the competitive baselines for speaker extraction and diarization tasks on LibriMix and SparseLibriMix datasets. We further validate the diarization performance on CALLHOME, a dataset based on real recordings, and experimental results indicate that our model surpasses recently proposed approaches. 
\end{abstract}


\section{Introduction}
\IEEEPARstart{E}{ach} speaker has unique voice characteristics. 
Speaker extraction seeks to extract speech from the target speaker with reference to his/her voiceprint \cite{zmolikova2023neural}, while speaker diarization analyzes the audio signal to determine `who spoke when' \cite{park2022review}.
As important speech processing front-ends, they are widely used in various real-world speech applications, such as speaker verification \cite{rao19_interspeech,xu2021sv} and speech recognition \cite{zmolikova2017speaker,marc2018seforasr,radford2023robust,mansfield21_interspeech}.

In a multi-talker scenario, the human brain processes incoming speech signals by performing speaker extraction and diarization simultaneously.
Prior studies typically treated speaker extraction~\cite{ge20_interspeech, xu2020spex, liu2023x} and diarization~\cite{Fujita2019eend,horiguchi2020end,Medennikov2020tsvad,jiang2023prompt,chen2023attention} as two independent tasks. 
However, in practical applications, such as meeting analysis and mobile recording applications, solely relying on the results from one task is insufficient since it can only provide speaking boundaries (timestamp information) or clean speech (content information). 
Associating extracted speech and speaker labels offers more meaningful and interpretable results as it provides a comprehensive view of `who spoke what and when' \cite{chen2020continuous}. %
Moreover, this association provides clean speech signals so that it can further facilitate the speaker-attributed automatic speech recognition task, which aims to answer `who spoke what'~\cite{shafey2019joint,kanda2020joint,kanda2021minimum}.

Given their shared objective of disentangling speakers from a speech mixture, we believe that speaker extraction and speaker diarization complement one another. 
On the one hand, speaker extraction leads to a clean speech signal, therefore enhancing the effectiveness of speaker activity detection in speaker diarization.
On the other hand, the information regarding the presence or absence of the target speaker resulting from speaker diarization is helpful for speaker extraction.

However, integrating speaker extraction and speaker diarization is challenging because of several critical differences between these tasks.
First, their outputs are presented in very different ways.
A speaker extraction system \cite{ge20_interspeech,xu2020spex,liu2023x} typically generates a single speaker's voice. 
In contrast, a speaker diarization system \cite{Fujita2019eend,horiguchi2020end,Medennikov2020tsvad} seeks to annotate the speech activities for all speakers.
Second, their application scenarios are significantly different. 
A speaker extraction system is usually optimized for highly overlapped speech, e.g. with a speaker overlapping ratio of near 100\% \cite{cosentino2020librimix}.
Conversely, speaker diarization studies mainly handle situations with sparsely overlapped speech, as the speaker overlap ratio is around 20\% in daily conversations and meetings \cite{cetin06_interspeech,barker18_interspeech}.
We define these two differences as the `output inconsistency' and the `scenario mismatch' issues, which need to be addressed when integrating them.

Most of the previous studies dealt with speaker extraction and speaker diarization either independently \cite{horiguchi2020end,Medennikov2020tsvad,ge20_interspeech,liu2023x} or jointly under some controlled conditions \cite{boeddeker2023ts,ueda2022eend}, such as with a fixed number of speakers or a pre-defined overlap ratio.
To date, no single-model approach has been proposed to perform speaker extraction and speaker diarization for speech mixture\modify{s} with varying overlap ratios and variable number of speakers. 
This prompts us to look into effectively incorporating speaker extraction and speaker diarization to benefit from their interaction.

With this motivation, we propose the Universal Speaker Extraction and Diarization model (USED) to address output inconsistency and scenario mismatch. 
The term `universal' is defined by two aspects: (a) the ability to perform for a variable number of speakers, and (b) the ability to process speech mixture\modify{s} with an arbitrary overlap ratio.

To this end, we design the model by considering two aspects: data ingress and data egress. 
From a data ingress perspective, we design a novel embedding assignment module driven by speech references to mitigate output inconsistency naturally. 
This module yields two advantages.
Firstly, it enables the generation of outputs for a variable number of speakers according to the speech references, thereby ensuring consistency across tasks.
Secondly, it helps the model align speaker extraction and speaker diarization outputs to avoid the output permutation problem \cite{yu2017permutation}. 
From a data egress perspective, we extend the scope of overlap ratio by a multi-task interaction module to alleviate the scenario mismatch between speaker extraction and speaker diarization. 
Specifically, the multi-task interaction module leverages speaker diarization results to mitigate background noise of the extracted speech and enhance the optimization of scenario-aware differentiated (SAD) loss \cite{zexu2022usev} within our USED.
Simultaneously, speaker extraction produces a cleaner speech signal to assist in the prediction of speaker diarization, particularly for the overlap part\modify{s}.

The contributions of this paper can be summarized as follows:
\begin{itemize}
    \item We propose a universal solution, i.e. USED, that performs the speaker diarization and extraction jointly. The solution deals with speech mixture\modify{s} with an arbitrary number of speakers and diverse overlap ratios.
    \item We design an embedding assignment module to ensure consistency in the number and the order of model outputs. This aids in supporting the generation of results for a variable number of speakers and enhances the model's robustness.
    \item We design a novel multi-task interaction module with a scenario-aware differentiated loss that allows the diarization output to control whether the target speech is silenced, thereby ensuring temporal overlap consistency between the diarization and extraction outputs.
    \item Our USED model outperforms competitive baseline systems of speaker extraction and speaker diarization on both the LibriMix and SparseLibriMix datasets. Additionally, we assess the diarization performance on the CALLHOME dataset, which consists of real recordings. The experimental results show that our model performs better with significantly less pre-training data.
\end{itemize} 

The rest of the paper is organized as follows.
Section \ref{sec:relatedwork} discusses the related work. Section \ref{sec:used0} describes the proposed method. We introduce the experimental setup in Section \ref{sec:exp_setup}. Section \ref{sec:exp_result} presents the experimental results and analysis. Section \ref{sec:conclu} concludes the study.

\section{Related Work}
\label{sec:relatedwork}
\subsection{Speech Separation and Speaker Extraction}
Recent studies on blind speech separation have made significant progress, with the development of advanced models such as Conv-TasNet~\cite{luo2019tasnet}, Dual-Path Recurrent Neural Network (DPRNN)~\cite{yi2020dprnn}, Band-Split Recurrent Neural Network (BSRNN)~\cite{luo2023music} and efficient training algorithms such as Permutation Invariant Training (PIT)~\cite{yu2017permutation}.
However, speech separation methods suffer from the global permutation ambiguity problem \cite{xu2020spex}, which makes it hard to process long utterances.
Moreover, it requires prior knowledge or estimation of the number of speakers in the speech mixture in advance, which is hard to get in real-world applications. 
Unlike blind speech separation that seeks to separate all speakers, speaker extraction only extracts speech for one target speaker at a time. 
In general, speaker extraction methods are classified into frequency-domain methods, such as SpeakerBeam \cite{zk2017spkbeam1,dm2019spkbeam2,zk2019spkbeam3} and VoiceFilter \cite{Wang2019VoiceFilterTV}, and time-domain methods, e.g. SpEx+ \cite{ge20_interspeech} and X-SepFormer \cite{liu2023x}.
Time-domain approaches can naturally avoid the phase estimation problem, which exists in frequency-domain approaches \cite{ge20_interspeech}.
Therefore, we develop the USED model based on the time-domain approach, SpEx+.

It is common that speech separation and speaker extraction systems in the literature are optimized for highly overlapped speech mixtures \cite{luo2019tasnet,ge20_interspeech,liu2023x}.
However, the overlap ratio in daily conversations and meetings is typically around 20\% \cite{cetin06_interspeech,barker18_interspeech}, leading to a scenario mismatch between training on highly overlapped speech and evaluating on sparsely overlapped speech.
To address such mismatch, speaker extraction for sparsely overlapped speech was recently studied \cite{borsdorf21_interspeech,Zhang2020XTaSNetRA,zexu2022usev}.
It was suggested to minimize speech power when the target speaker is absent, while maximizing the signal-to-distortion ratio (SDR) or scale-invariant signal-to-distortion ratio (SI-SDR) \cite{roux2019sisnr} when the target speaker is present.
This introduces another challenge since it is hard to balance the two scenario-dependent objectives, i.e., SI-SDR and power, during training. 
We will study a multi-task interaction module in the USED model to overcome this challenge. 

\subsection{Speaker Diarization}
Traditional clustering-based diarization methods \cite{shum2013unsupervised,shell2014diar,Mohammed2014sd,8462628,wang2024overview} typically comprise multiple components that are trained separately.
These methods implicitly assume that each speech segment is only from one speaker, which fails when multiple people speak at the same time.
Two mainstream studies are attempting to solve this problem, i.e. end-to-end neural diarization (EEND)~\cite{Fujita2019eend,horiguchi2020end}, and target speaker voice activity detection (TS-VAD)~\cite{Medennikov2020tsvad}.
EEND seeks to minimize the diarization error directly with permutation-free objectives \cite{Fujita2019eend}, while TS-VAD uses the speaker embedding from the clustering results as the reference to detect the speaking status of each person.
These methods have obtained good performance and set the stage for our study.

\subsection{Interaction Between Speech Separation and Diarization}
There were studies on the interaction between speech separation and diarization~\cite{Fujita2019eend,wang2021scenario} by training speech processing models with multiple tasks.
EEND-SS \cite{ueda2022eend} is a joint end-to-end framework for speaker diarization and separation. 
It utilizes the bottleneck feature of separation as one of the inputs for diarization and is optimized using multi-task learning. 
This model faces challenges related to permutation and processing very long sentences. 
In addition, the fusion technique of EEND-SS is only applied during inference by using diarization probability directly. 
In TS-SEP \cite{boeddeker2023ts}, it was proposed to pre-train a model with TS-VAD objective, and subsequently finetune the model to predict the mask for separation. 
In other words, TS-SEP is a two-stage framework that focuses on the speech separation task. 
Unlike EEND-SS and TS-SEP, our USED is a universal model that can concurrently handle both tasks and address their output inconsistency and scenario mismatch issues.

Two \modify{recent} works \cite{kalda2024pixit,bando2024neural} have explored the joint training of speech separation and speaker diarization based on unsupervised speech separation approaches.
PixIT \cite{kalda2024pixit} employs PIT loss to integrate the speaker diarization task with MixIT \cite{wisdom2020unsupervised} and draws on the \textit{best-of-both-worlds} framework \cite{kinoshita2021advances,9414333} to stitch different segments based on the speaker diarization results.
On the other hand, Neural FCASA \cite{bando2024neural} combines speaker diarization with neural FCA \cite{9506855,bando22_interspeech}, eliminating the need for a separate speaker diarization step to mask source activities.
Both approaches jointly address the two tasks under the premise of unsupervised speech separation, focusing on long-form audio and optimizing the final separation results in terms of diarization error rate (DER) and word error rate (WER).
In comparison, our USED model emphasizes the performance of supervised speech separation and speaker diarization across different overlap ratios and scenarios.

\section{Universal Speaker Extraction and Diarization}
\label{sec:used0}

\subsection{Task Definition}
Let $x$ be a multi-talker speech mixture with varying overlap ratio from 0\% to 100 \%, which consists of the speech signals from $l$ speakers and \modify{a} noise signal $n$. 
The speech mixture can be represented as,
\begin{equation}
\label{equ:mixture}
    x = \sum^{l}_{i=1} c_i + n
\end{equation}
where $l>1$, and $c_i$ denotes the speech signal of speaker $i$.
Each speaker $i$ has a corresponding speech reference, denoted as $z_i$, which is estimated from \modify{the} speech mixture or provided as pre-enrolled information. %
\modify{Note that no normalization or noise reduction techniques are applied to the raw speech signals.}

Given a speech mixture and speech references, the goal of the USED model is to generate separated speech signals and speaker activity information concurrently for a variable number of speakers, ranging from $1$ to $l$.
The number of speakers corresponds to the available speech references.
This is motivated by scenarios where obtaining speech references for all speakers is challenging, or the target speakers are only a subset of all speakers.
During inference, if there is only one target speaker, the USED model outputs a speech signal and diarization result for that speaker in a similar way to a speaker extraction model~\cite{ge20_interspeech}.
However, if there are $l$ target speakers, the model would output multiple speech signals and \modify{a} diarization results concurrently, one for each speaker, just like a speech separation model does~\cite{luo2019tasnet}.

\label{sec:used}
\subsection{System Overview}
\label{sec:overview}

\begin{figure*}[htb]
  \centering
  \centerline{\includegraphics[width=0.95\linewidth]{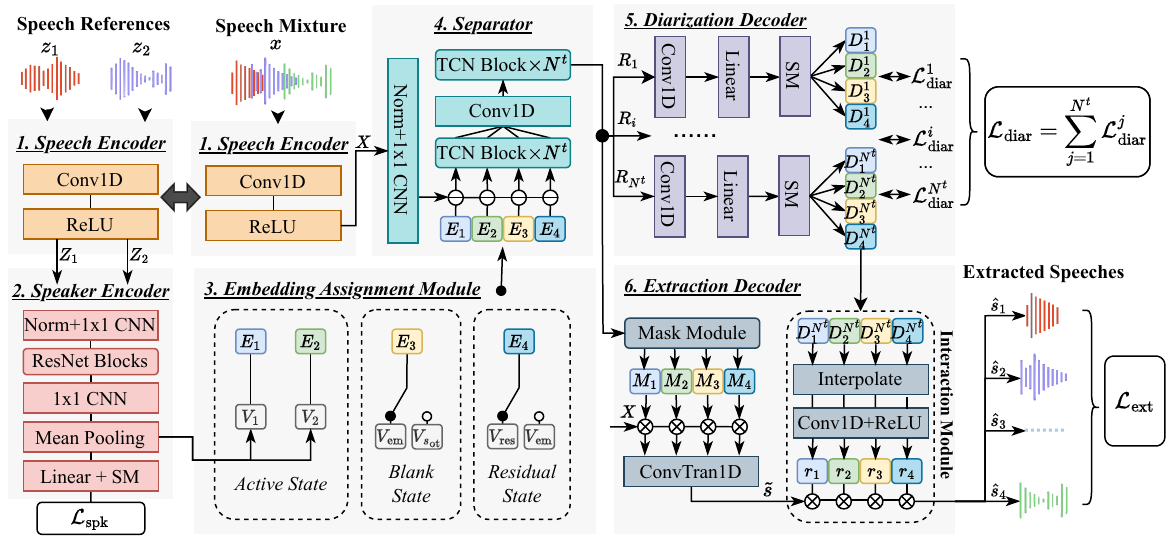}}

\caption{ 
The overall training and inference framework of our proposed USED model. 
It comprises a speech encoder, a speaker encoder, an embedding assignment module, a separator, an extraction decoder and a diarization decoder.
The USED model generates both \modify{the} extracted speech and speaker \modify{diarization} results by leveraging the speech mixture and \modify{the speech} reference\modify{s} as input.
The symbols $\otimes$ and $\ominus$ refer to element-wise multiplication and frame-wise concatenation, respectively.
The TCN block and mask module are denoted by rounded rectangles, which are described in detail in Fig. \ref{fig.modules} \modify{and} introduced later.
Different colours are used to distinguish network layers from different components.
The \textit{active}, \textit{blank} and \textit{residual} state\modify{s} are assigned by Algorithm \ref{alg:1}.
}
\label{fig.USED}
\vspace{-10pt}
\end{figure*}

As illustrated in Fig. \ref{fig.USED}, the USED model comprises six components: a speech encoder, a speaker encoder, an embedding processing module, a separator, an extraction decoder, and a diarization decoder.

The speaker encoder converts the available speech references into speaker embeddings, that are the feature representations of the target speakers.
Meanwhile, the speech encoder generates spectrum or spectrum-like feature representations from the speech mixture $x$.

Unlike TS-VAD \cite{Medennikov2020tsvad}, which generates diarization results for all speakers according to speaker embeddings and \modify{the} speech mixture, we propose an embedding assignment module to enable custom control over the number of model's outputs.
The embedding assignment module prepares the input embeddings with three different states, \textit{active}, \textit{blank}, and \textit{residual}, based on the speaker embeddings, which are responsible for generating results accordingly.
The \textit{active} state is used as a position marker for the expected speakers. 
The \textit{blank} state serves as a position marker to guide the network in outputting a silent segment.
The \textit{residual} state marks the residual unexpected speakers.
The \textit{blank} state is motivated by Connectionist Temporal Classification (CTC) \cite{graves2006connectionist}, which integrates blank characters to manage pauses in speech recognition and gaps between characters in Optical Character Recognition (OCR).
The embedding with \textit{blank} state is designed to clearly distinguish the outputs from the expected and unexpected speakers.
The reason for this design is that the output from an unexpected speaker is usually a silent segment.
With the help of the embedding assignment module, the USED model naturally avoids the output inconsistency stemming from the integration of speaker extraction and diarization.

The separator then aims to separate speakers at the representation level from a speech mixture and embeddings from the embedding assignment module.
Finally, diarization and extraction decoders predict the corresponding results based on the output of the separator.

The USED model differs from typical speaker extraction and speaker diarization systems \cite{Fujita2019eend,ge20_interspeech} in two key aspects.
Firstly, it outputs speech signals for an arbitrary number of speakers, ranging from $1$ to $l$, with the help of the embedding assignment module.
Secondly, it seeks to effectively deal with speech mixture\modify{s} with a wide range of overlapping ratios.

\subsection{Speech Encoder}
Analogous to a frequency analyzer, the speech encoder aims to generate spectrum-like frame-based embedding sequences from \modify{the} speech mixture and speech reference\modify{s}. 
Inspired by SpEx+\cite{ge20_interspeech}, our approach leverages a multi-scale speech encoder.

Our multi-scale speech encoder is structured as a combination of three one-dimensional Convolutional Neural Networks (Conv1D), each utilizing distinct kernel sizes to represent different scales.
Subsequent to the convolutional layers, the activation function applied is the Rectified Linear Unit (ReLU).

The speech encoder produces spectrum-like frame-based embedding sequences $X$ and $Z_i$ for 3 scales as defined in \cite{ge20_interspeech} from the speech mixture $x$ and the $i$-th speech reference $z_i$ as
\begin{equation}
    X = [X_1, X_2, X_3] = e(x) \in \mathbb{R}^{3 \times T\times C}
\end{equation}
\begin{equation}
    Z_i = [Z_{i1}, Z_{i2}, Z_{i3}] = e(z_i) \in \mathbb{R}^{3 \times T\times C}
\end{equation}
where the function $e(\cdot)$ denotes the operation performed by the multi-scale speech encoder.
The set $Z = \{Z_1, ..., Z_{l}\}$ represents the embedding sequences for $l$ speakers.
The input and output channel sizes of the three Conv1Ds are one and $C$, respectively.
The kernel sizes of \modify{the} three Conv1Ds are denoted as $L^{e}_1$, $L^{e}_2$ and $L^{e}_3$, while they have a common stride of $\frac{L^{e}_1}{2}$.
The outputs $X$ and $Z_i$ have a shape of $3 \times T \times C$, where $T$ is the number of frames. 
For simplicity, Fig. \ref{fig.USED} does not explicitly depict the multi-scale structure.
\begin{figure}%
    \centering
    \subfloat[\centering ResNet Block\label{fig.resnet}]{{\includegraphics[width=0.3\linewidth]{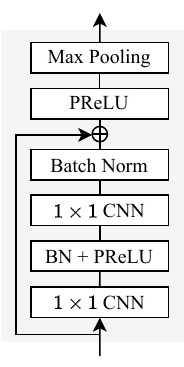} }}
    \subfloat[\centering TCN Layer\label{fig.tcn}]{{\includegraphics[width=0.3\linewidth]{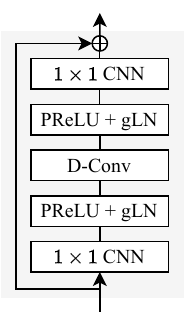} }}%
    \caption{\label{fig.modules}\modify{M}odel structure for \modify{a} ResNet block and \modify{a} TCN layer. BN, PReLU, gLN and D-Conv are batch normalization, parametric ReLU, global layer normalization and dilated depth-wise separable convolution. The symbol $\oplus$ refers to element-wise addition.}
\vspace{-10pt}
\end{figure}
\begin{algorithm2e}
\small
\caption{Assigning States for Embeddings}\label{alg:1}
\KwData{ \\
    $k$: Maximum number of speakers the model can process \\
    $S_{\text{mix\_spk}}$: Set of speaker IDs for the speech mixture\\
    $S_{\text{all\_spk}}$: Set of speaker IDs for the whole dataset\\
    $V_{\text{em}}$: Empty embedding \\
    $V_{\text{res}}$: Residual embedding \\
    $p_{\text{sel}}$: Probability for selecting present speaker \\
    $p_{\text{em}}$: Probability to use empty embedding as input \\
    $p_{\text{res}}$: Probability to use residual embedding as input \\
}
\tcp{List of embeddings with active and blank state}
$I_{active}, I_{blank} \gets [$ $], [$ $]$\;
\tcp{Number of embeddings}
$count \gets 0$\;
\tcc{The First Step for Active State}
\For{$s_{\text{mix}} \in S_{\text{mix\_spk}}$}{
    \If{$s_{\text{mix}}$ is a present speaker}{
    $p \gets $ random\_uniform$(0, 1)$\;
        \tcp{If true, then marking embedding as \textit{active} state}
        \If{$p>p_{\text{sel}}$}
        {
            $E_{count} \gets V_{s_{\text{mix}}}$\;
            $I_{active}$.add($E_{count}$)\;
            $count \gets count + 1$\;
        }
    }
}

\tcc{The Second Step for Blank State}
\While{$count \leq k$}{
    $p \gets $ random\_uniform$(0, 1)$;

    \eIf{$p > p_{\text{em}}$}{
        \tcp{Randomly select one speaker from the set of speakers that includes $S_{\text{all\_spk}}$ but excluding $S_{\text{mix\_spk}}$}
        $s_{\text{ot}} \gets $random\_sample$(S_{\text{all\_spk}} \setminus S_{\text{mix\_spk}})$\;
        $E_{count} \gets V_{s_{\text{ot}}}$\;
    }{
        $E_{count} \gets V_{\text{em}}$\;
    }
    $I_{blank}$.add($E_{count}$)\;
    $count \gets count + 1$\;
}
\tcc{The Third Step for Residual State}
$p \gets $ random\_uniform$(0, 1)$\;
\eIf{$p > p_{\text{res}}$ or $ I_{active}.\text{length} \neq$ number of present speakers}{
    $E_{k+1} \gets V_{\text{res}}$\;
}{
    $E_{k+1} \gets V_{\text{em}}$\;
}
\textbf{Output:} $I_{active}$, $I_{blank}$, $E_{k+1}$
\end{algorithm2e}

\subsection{Speaker Encoder}
The speaker encoder is designed to extract speaker embedding\modify{s} from the corresponding enrolled speech. Specifically, we employ a speaker encoder to extract speaker embeddings, denoted as $V_1, ..., V_l$, from the embedding sequences $Z$.
These speaker embeddings capture the distinctive characteristics of each speaker. 
After layer normalization (LN), we leverage a CNN layer with a kernel size of $1 \times 1$ ($1 \times 1$ CNN) on the embedding sequence $Z_i$ for speaker $i$.
This is subsequently followed by $N^r$ residual network (ResNet) blocks \cite{he2016deep}.
The input and output channel sizes of this $1 \times 1$ CNN are $3C$ and $C$, respectively.
Finally, an additional $1 \times 1$ CNN, combined with a mean pooling operation, generates a speaker embedding $V_i$ for speaker $i$ with an embedding dimension $D_{\text{spk}}$.
After that, the speaker encoder predicts speaker identities, denoted as $\hat{y}^{\text{spk}}_i$, through a linear layer followed by a softmax (SM) activation function for speaker $i$.

The detailed architecture of the ResNet we used is shown in Fig. \ref{fig.resnet}. \modify{A} ResNet block comprises two $1 \times 1$ CNNs and a max-pooling layer.
Each $1 \times 1$ CNN within the ResNet block is accompanied by batch normalization (BN) and parametric ReLU (PReLU) for normalization and non-linear transformation.
A skip connection adds the input to the output obtained after the second BN layer. 
Lastly, the max-pooling layer is responsible for adjusting the length of the output embedding.

We choose ResNet \modify{\cite{he2016deep}} because it has become a prominent architecture in speaker verification systems \cite{chung2020defence, zhou2021resnext, wang2023wespeaker, zeinali2019but} and is widely adopted in speaker extraction models \cite{ge20_interspeech, wang24fa_interspeech}.
Its key advantage lies in the residual connections, which link frame-level layers and mitigate the vanishing gradient problem, facilitating faster convergence during backpropagation \modify{\cite{he2016deep}}.
These residual connections also enable deeper neural network construction, allowing ResNet-based models to outperform regular CNNs.
As a result, using ResNet as the backbone for speaker encoders leads to more robust and higher-quality speaker embeddings, making it a natural choice for this task.

\subsection{Embedding Assignment Module}
\label{speaker_shuff}
The embedding assignment module aims to prepare embeddings with different states for the separator.
We set embeddings with different states: \textit{active}, \textit{blank}, and \textit{residual}.
The details of the embedding assignment module are shown in Algorithm~\ref{alg:1}.

\subsubsection{Active State} The embedding with \textit{active} state is designed to guide the network to generate outputs for the expected speakers. %
Based on the embedding with \textit{active} state, the whole speaker extraction and diarization process is driven by the speaker encoder that takes the speech references of some present speakers as input. 
Specifically, to enable our model to achieve the correct output for an arbitrary number of speakers, i.e., from 1 to $l$, we randomly assign the \textit{active} state to the embeddings of present speakers by a hyperparameter $p_{\text{sel}}$ representing probability, as shown in the first step of Algorithm~\ref{alg:1}.

\subsubsection{Blank State} The embeddings with \textit{blank} state are primarily responsible for driving the network to produce silent segments for the unexpected speakers, which is critical in ensuring our network's robustness. 
Unlike \textit{active} state, which requires a unique embedding from the speaker encoder, the embeddings with \textit{blank} state alternate between two embeddings via a threshold $p_{\text{em}}$ to guide the network: an inactive speaker embedding and a learnable empty embedding. 
The inactive speaker embedding is derived from a speaker who is not present in the speech mixture.
This design ensures that our USED model can customize the number of outputs by inputting the learnable empty embedding during the inference stage.

\subsubsection{Residual State} As an additional state, the targets corresponding to the embedding with \textit{residual} state are the cumulative outputs for speakers that are present but are not selected at the first step.
This design allows our USED model to explore the interaction among all speakers in the speech mixture. 
Specifically, we also employ a learnable embedding $V_{\text{res}}$ to teach the network to learn the corresponding cumulative outputs for the residuals. 
When speaker embeddings of all speakers have been considered as the \textit{active} state, the target corresponding to \textit{residual} state should be left without any speakers. 
In such cases, the choice between $V_{\text{res}}$ and $V_{\text{em}}$ is determined through a specified threshold $p_{\text{res}}$. 
Our preliminary experiments demonstrate that this strategy enhances robustness during the inference phase.

Overall, our USED model has $(k+1)$ input embeddings, where the first $k$ embeddings with \textit{active} and \textit{blank} states support the predictions of at most $k$ speakers and the $(k+1)$-th embedding $E_{k+1}$ is with the \textit{residual} state.
Finally, we shuffle all the embeddings except $E_{k+1}$ to guarantee the model's insensitivity to the order of embeddings.
Through the designed embedding processing mechanism, our USED model can predict the outputs from a custom number of speakers. 
Additionally, the model's robustness is enhanced by employing a random selection strategy for speakers when assigning \textit{blank} state.

\subsection{Separator}
The separator is designed to separate speakers based on the embeddings from the embedding assignment module.
As shown in Fig. \ref{fig.USED}, the embedding sequence $X$ firstly undergoes a series of operations, including a layer normalization followed by $1 \times 1$ CNN layer, wherein the input channel size is set to $3C$ and the output channel size is $C$.
Then, we introduce $N^t$ Temporal Convolutional Network (TCN) blocks \cite{luo2019tasnet} to generate output representations based on the embeddings and speech mixture.

\modify{We use TCN as the backbone of our model because they have been widely adopted in speech separation and speaker extraction tasks\cite{luo2019tasnet,ge20_interspeech}.
The USED model is built on the SpEx+ model, which also utilizes TCN.
This choice ensures consistency and enhances performance.
In addition, TCN is commonly used for time-domain approaches, avoiding the phase estimation issues found in time-frequency models.
Finally, compared to methods like DPRNN \cite{yi2020dprnn} and BSRNN \cite{luo2023music}, Conv-TasNet \cite{luo2019tasnet} offers lower computational complexity, reduced memory requirements, and faster training and inference.}

Representations are then concatenated and down-sampled via a Conv1D layer with input channel size $(k + 1) \times C$ and output channel size $C$.
Finally, the processed representation is forwarded to another $N^t$ TCN blocks to generate representations $\{R_1, ..., R_{_{N^t}}\}$.

Each TCN block comprises a set of $N^b$ TCN layers.
\modify{The} dilated depth-wise separable convolution (D-Conv) shown in Fig. \ref{fig.tcn} has an exponential growth dilation factor $2^b$, where $b \in \{0, ..., N^b - 1\}$.
In the initial TCN layer within each block, the input channel size for the $1 \times 1$ CNN is $D_{\text{spk}} + C$.

\subsection{Diarization Decoder}
The diarization decoder aims to predict the frame-level activity probabilities for each speaker from the $N^t$ TCN blocks' outputs $\{R_1, ..., R_{_{N^t}}\}$ in the separator.
Because speaker extraction and diarization require different time resolutions, we employ a Conv1D layer to down-sample the outputs from the TCN block, with input and output channel sizes of $C$, a kernel size of $L^{\text{diar}}$, and a stride of $L^{\text{diar}} / 2$.
Subsequently, we use a linear layer followed by a softmax operation to compute the probabilities, denoted as $D^j = \{D^{j}_{1}, ..., D^{j}_{k+1}\}$, which represent the speech activities of the corresponding embeddings.
Finally, we get a set of diarization results from each TCN block, denoted as $\{D^{1}, ..., D^{N^t}\}$.
During inference, we only use the diarization result $D^{N^t}$ from the final TCN block as the results to calculate the evaluation metric.

\subsection{Extraction Decoder}
The extraction decoder is designed to estimate masks for each speaker and reconstruct the corresponding signals. \modify{T}he mask module of \modify{the} extraction decoder comprises a Conv1D 
\modify{layer} and ReLU activation to generate masks $\{M_1, ..., M_{k + 1}\}$, where \modify{$M_i \in \mathbb{R}^{T \times C}$}.
We then obtain the modulated responses $i$ by element-wise multiplication of the mask $M_i$ and the representations $X$ from \modify{the} speech mixture.

For signal reconstruction, we employ a multi-scale decoder like speech encoder, which consists of three transposed convolutional networks (ConvTran1D), each with a kernel size of $L^{\text{e}}_1$, $L^{\text{e}}_2$ and $L^{\text{e}}_3$, and a common stride, $L^{\text{e}}_1 / 2$. 
It's used to reconstruct the time-domain signals, $\tilde{s} = \{\tilde{s}_1, ..., \tilde{s}_{k + 1}\}$, where $\tilde{s}_i = \{\tilde{s}^1_i, \tilde{s}^2_i, \tilde{s}^3_i\}$ is \modify{obtained} from three ConvTran1D layers.

As the USED model generates both the extracted speech and speaker diarization results, we propose an interaction module (IM) to use speaker diarization results to refine the generated waveforms.
To avoid the gradient from the extraction task \modify{having an effect} on the diarization task, we first create a clone of $D^{N^t} = \{D^{N^t}_{1}, ..., D^{N^t}_{(k+1)}\}$ without gradient. 
Subsequently, we employ interpolation techniques on the diarization output probabilities to ensure length-based uniformity between the two tasks.
Following this alignment, a layer comprising a Conv1D layer and ReLU activation is employed to further process the output probabilities and obtain outputs $r_i$, ensuring that the values at each step are not constrained to be less than 1.
The Conv1D layer possesses an input channel size of 1, an output channel size of 1, and a kernel size of $L^{\text{im}}$.

Finally, the model generates the extracted speeches, $\{\hat{s}_1, ..., \hat{s}_{k + 1}\}$, through element-wise multiplication of $\{r_1, ..., r_{k + 1}\}$ and \modify{the output} $\tilde{s}$ from \modify{the} extraction module separately as
\begin{equation}
    \hat{s}_i = \{ \hat{s}^1_i, \hat{s}^2_i, \hat{s}^3_i \}
\end{equation}
where
\begin{equation}
    \hat{s}^j_i = \tilde{s}^j_i \otimes r_i
\end{equation}

It should be noted that, during inference, only $\tilde{s}^1_1, ..., \tilde{s}^1_k$ originating from the first ConvTran1D are utilized for the evaluation of metrics.

\subsection{Loss Function}
\subsubsection{Overall Loss Function}
The loss of the USED model can be formulated as follows:
\begin{equation}
    \label{eq.overall}
    \mathcal{L} = \lambda_1 \mathcal{L}_{\text{ext}} + \lambda_2 \mathcal{L}_{\text{diar}} + \lambda_3 \mathcal{L}_{\text{spk}}
\end{equation}
where $\lambda_1$, $\lambda_2$ and $\lambda_3$ are the weights for speaker extraction loss, speaker diarization loss, and speaker classification loss, respectively.
\begin{figure}[t]
  \centering
  \centerline{\includegraphics[width=6cm]{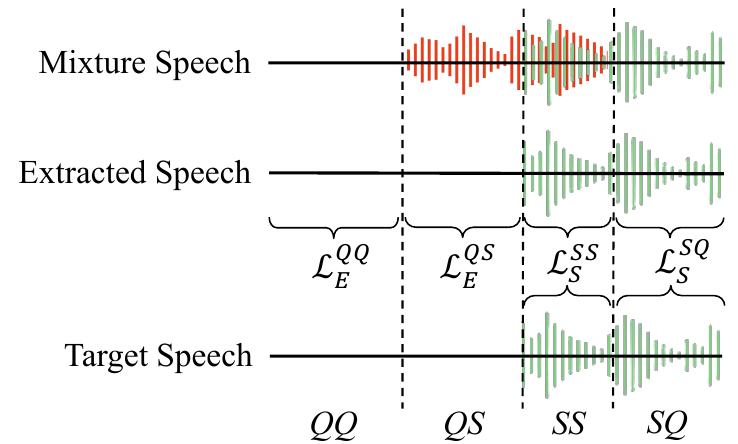}}
\caption{Illustration of scenario-aware differentiated loss. The speech mixture is segmented according to the four scenarios (\textit{QQ}, \textit{QS}, \textit{SS} and \textit{SQ}).}
\label{fig.loss}
\vspace{-10pt}
\end{figure}
\subsubsection{Speaker Extraction}
\label{sec:loss_se}
We adopt the SAD loss proposed in \cite{zexu2022usev} to address sparsely overlapped speech as shown in Fig. \ref{fig.loss}.
Scenarios are classified into four distinct classes: \textit{QQ}, \textit{QS}, \textit{SS}, and \textit{SQ}.
In the \textit{QQ} scenario, both the target speaker and interference speakers are in a \textit{quiet} state.
The \textit{QS} scenario represents a situation where the target speaker is \textit{quiet} while the interference speakers are \textit{speaking}.
Conversely, in the \textit{SQ} scenario, the target speaker is \textit{speaking} while the interference speakers are \textit{quiet}.
Lastly, the \textit{SS} scenario corresponds to both the target speaker and interference speakers being in a \textit{speaking} state.
For \textit{QS} and \textit{QQ}, we use power as the objective for extracted speech $\hat{s}$,
\begin{equation}
    \mathcal{L}_{\text{E}} = \sum^3_{j=1} 10\mu_j \log_{10}(\frac{\Vert \hat{s}^j\Vert ^ 2}{T_{\text{se}}}  + \epsilon)
\end{equation}
where $T_{\text{se}}$ is the duration of $\hat{s}^j$ in seconds, $[\mu_1, \mu_2, \mu_3]$ are the weights for different ConvTran1D layers and $\epsilon$ is set to $10^{-6}$.
For \textit{SS} and \textit{SQ}, we compute the SI-SDR loss between the target speech $s$ and extracted speech $\hat{s}$ as $\mathcal{L}_\text{S}$,

\begin{equation}
    \label{eq:sisnr}
    \mathcal{L}_{\text{S}} = \sum^3_{j=1} -10\mu_j \log_{10} \frac{\Vert \frac{<\hat{s}^j,s>s}{\Vert s \Vert^2+\epsilon} \Vert^2}{\Vert \hat{s}^j -  \frac{<\hat{s}^j,s>s}{\Vert s \Vert^2+\epsilon} \Vert^2+\epsilon} + \epsilon
\end{equation}

SAD loss is used to calculate the loss between the extracted speeches $\{\hat{s}_1, ..,  \hat{s}_{k + 1}\}$ generated from the multi-task interaction module and target speeches $\{s_1, ..., s_{k+1}\}$.
For clarity, we assume the first $m$ target speeches correspond to the embeddings with \textit{active} state and are the same as $\{c_1, ..., c_m\}$ in Eq. (\ref{equ:mixture}).
$\{s_{m+1}, ..., s_{k}\}$ are zero vectors which present silence as the target of embeddings with \textit{blank} state.
$s_{k+1}$ is the target speech for residual prediction.
In practice, the order is shuffled as introduced in section \ref{speaker_shuff}.
Each pair of extracted speech and target speech are firstly segmented according to the four scenarios.
Finally, the total speaker extraction loss, i.e. $\mathcal{L}_{\text{ext}}$, is as follows:

\begin{equation}
\label{eq:sadloss}
    \mathcal{L}_{\text{ext}} = \frac{1}{k + 1} (\sum_{i=1}^{k + 1} \alpha \mathcal{L}_{\text{E}_i}^{QQ} + \beta \mathcal{L}_{\text{E}_i}^{QS} + \gamma \mathcal{L}_{\text{S}_i}^{SS} + \delta \mathcal{L}_{\text{S}_i}^{SQ})
\end{equation}
where $k + 1$ is the total number of outputs and $\alpha$, $\beta$, $\gamma$ and $\delta$ are the weights for different scenarios.

\subsubsection{Speaker Diarization}
To optimize the model for the diarization task, we apply the binary cross-entropy loss (BCE), denoted as $\mathcal{L}_{\text{diar}}^j$ on the output of \modify{the} $N^t$ TCN blocks,

\begin{equation}
    \mathcal{L}_{\text{diar}}^j = \frac{1}{k + 1} \sum_{i=1}^{k + 1} BCE(y^{\text{diar}}_i, D_{i}^j)
\end{equation}
where $y^{\text{diar}}_{i}$ is the ground-truth label for the $i$-th output. Then, the entire speaker diarization loss can be represented as:

\begin{equation}
    \mathcal{L}_{\text{diar}} = \sum_{j=1}^{N^t} \mathcal{L}^{j}_{\text{diar}}
\end{equation}

\subsubsection{Speaker Classification}
The speaker encoder is optimized through a cross-entropy loss (CE), denoted as $\mathcal{L}_{\text{spk}}$, for speaker classification during the training process, as
\begin{equation}
    \mathcal{L}_{\text{spk}} = \frac{1}{l} \sum_{i=1}^{l} CE(y^{\text{spk}}_i, \hat{y}^{\text{spk}}_i)
\end{equation}
where $y^{\text{spk}}_i$ and $\hat{y}^{\text{spk}}_i$ are the ground truth and predicted speaker ids for speaker $i$, respectively.

\section{Experimental Setup}
\label{sec:exp_setup}
\subsection{Datasets}
We conduct comprehensive evaluations of our model using diverse datasets, including LibriMix and SparseLibriMix, which are simulated, and CALLHOME, which consists of real recordings.

\subsubsection{LibriMix}
The first dataset, known as LibriMix \cite{cosentino2020librimix}, generates samples derived from LibriSpeech \cite{panayotov2015librispeech} train-clean-100 and test-clean datasets, as well as WHAM! \cite{wichern2019wham}, all with a 16kHz sampling rate.
LibriMix has been widely used for speech separation \cite{li2023an, yang21c_interspeech}, speaker extraction \cite{lspex2022ge} and speaker diarization tasks\cite{yang21c_interspeech, ueda2022eend}.
We merge the Libri2Mix 100h and Libri3Mix 100h datasets, resulting in utterances that may contain either 2 or 3 speakers.

The models are evaluated on both min and max modes.
The min mode consists of highly overlapped speech segments, while a larger portion of the max mode consists of non-overlapped speech segments.
To ensure compatibility with the speaker extraction task, we have slightly adjusted the data split for training and validation \footnote{\url{https://github.com/msinanyildirim/USED-splits}}, as previously done in related work \cite{lspex2022ge}.
Specifically, we follow scripts \footnote{\url{https://github.com/gemengtju/L-SpEx/tree/main/data}} to prepare speech references for the Libri2Mix dataset and randomly select speech references for each speech mixture in the Libri3Mix dataset.

\subsubsection{SparseLibriMix}
The second dataset employed in our evaluation is the test set from SparseLibriMix \cite{cosentino2020librimix}, which utilizes WHAM! \cite{wichern2019wham} to simulate a noisy acoustic environment. 
This dataset covers more realistic mixture scenarios, varying from an overlap ratio of 0 to 1.0.
It contains 1,000 utterances which have 2 or 3 speakers.

\subsubsection{CALLHOME}
We use the CALLHOME dataset \cite{Mark_Martin_2001} to evaluate the diarization task with real recordings.
The CALLHOME dataset is partitioned into two parts based on the Kaldi recipe \footnote{\url{https://github.com/kaldi-asr/kaldi/tree/master/egs/callhome_diarization/v2}}. 
The first part, known as Part 1, is utilized for model adaptation, whereas the second part, referred to as Part 2, is used for evaluation.
The CALLHOME dataset consists of telephone-channel recordings with 8k sample rate.
Besides CALLHOME Part 1, we further add Libri2Mix 100h, Libri3Mix 100h, Libri2Mix 360h and Libri3Mix 360h, which are generated from LibriSpeech \cite{panayotov2015librispeech} train-clean-100 and train-clean-360, as the pre-training dataset.
In total, our dataset encompasses approximately 452 hours of data.
We then finetune the model only with CALLHOME Part 1 and evaluate on CALLHOME Part 2.
During inference, we follow the evaluation pipeline of TS-VAD to evaluate our model so that it does not require pre-enrolled speech.
Specifically, we extract the speech references based on the diarization result of spectral clustering and then do inference for USED models.
The speaker embedding for spectral clustering is extracted from the ECAPA-TDNN model \cite{desplanques2020ecapa} trained on VoxCeleb2~\cite{chung2018voxceleb2}~\footnote{\url{https://github.com/TaoRuijie/ECAPA-TDNN}}.

\subsection{Baseline Configuration}
Two of the baseline models, SpEx+ \cite{ge20_interspeech}, and TS-VAD \cite{Medennikov2020tsvad}, are implemented by ourselves.
During the training phase, we segment the utterances into chunks, each of which spans 4 seconds and is shifted every 2 seconds.

For the SpEx+ baseline, we use the same training configuration as the USED model described in section \ref{used_config}.
To align the model complexity with that of the USED model, we configure the SpEx+ baseline with a total of 8 TCN blocks so that it has a similar number of parameters to the USED model.
It is worth noting that the SpEx+ baseline can be regarded as a specialized version of the USED model, as it has only one embedding with \textit{active} state as input without including a diarization module and SAD loss.
Our metric for selecting the best checkpoint for the SpEx+ model is SI-SDR.

As for the TS-VAD baseline, we follow \cite{Medennikov2020tsvad} to design the model, with several notable modifications.
Firstly, we substitute the bidirectional LSTM in the original TS-VAD model with four transformer layers, with sine-cosine positional encoding as input.
The model architecture entails two layers dedicated to processing inputs for each speaker individually, followed by a Conv1D layer for down-sampling.
Another two layers are used to combine the information from all speakers. 
For the transformer, we set the embedding dimension to 384 and employ four attention heads.
Second, we use a pre-trained speaker encoder, ECAPA-TDNN \cite{desplanques2020ecapa}, to extract speaker embeddings.
This speaker encoder also takes on the role of replacing the CNN encoder to extract time-level representations from the speech mixture.
Finally, we use a similar strategy as stated in section \ref{speaker_shuff}, which randomly selects speaker embeddings from the whole dataset.
$p_{em}$ is set to be 0.3.
The model is optimized using Adam with batch size 64 for 40k steps.
The speech encoder is kept fixed for the first 4,000 steps, and the learning rate follows a polynomial decay schedule with an initial value of $2 \times 10^{-4}$, which has a warm-up phase for the first 10\% of steps.
The DER metric is our criterion for selecting the best checkpoint for the TS-VAD model.

\subsection{USED Model Configuration}
\label{used_config}
The maximum number of speakers $k$ is 3 for evaluation on the LibriMix and SparseLibriMix datasets and in the assessment on the CALLHOME dataset.
For the multi-scale speech encoder, $C$ is 256 and the kernel sizes,  $L^e_1$, $L^e_2$ and $L^e_3$, are 20, 80 and 160.
The number of ResNet blocks $N^r$ in the speaker encoder is set to 4, and the dimension of the speaker embedding $D_{\text{spk}}$ is 256.
Regarding the separator module, we set $N^t$ to 3 and $N^b$ to 8.
For the diarization decoder, the CNN layer has a kernel size $L^{\text{diar}}$ of 32 and stride $L^{\text{diar}} / 2$ of 16.
The mask module of the extraction decoder consists of a CNN with a kernel size of 1 and stride 1, and ReLU activation.
For the multi-task interaction module, the kernel size $L^{\text{im}}$ of the CNN layer is 16.
The loss weights for outputs from different ConvTran1D layers, $\mu_1$, $\mu_2$ and $\mu_3$ are 0.8, 0.1 and 0.1 by following \cite{ge20_interspeech}.
We set $0.001$ for $\alpha$ and $\beta$, and $1.0$ for $\gamma$ and $\delta$, which are the loss weights in Eq. (\ref{eq:sadloss}).
$\lambda_1$, $\lambda_2$, and $\lambda_3$ are set to 1.0.
$p_{\text{em}}$ is set to 0.3.
$p_{\text{sel}}$ and $p_{\text{res}}$ are set to 0.5 and 0.9, respectively, when \textit{residual} state is applied.
\modify{T}he USED model without \textit{residual} state and with $p_{sel} = 0$ \modify{is} call\modify{ed} as ``USED-F(ix)'' since it always gets speech references of present speakers during training.

We optimize the model with Adam on 4 GPUs for 100k steps, corresponding to approximately 11 epochs.
The utterances are split into chunks with a size of 4 seconds and a shift of 2 seconds.
We set the maximum tokens to 260k, which results in processing around 16 seconds of audio for each batch.
The learning rate is set to $1e-3$ for all experiments, which is warmed up for the first 10\% steps and decayed polynomially for the \modify{remaining} steps.
The overall loss value is our primary metric for selecting the best checkpoint for the USED model.

\subsection{Evaluation Metrics}

We use diarization error rate (DER (\%)), including speaker confusion (SC (\%)), false alarm (FA (\%)), and missed detection (MS (\%)) to evaluate the speaker diarization performance. 
Collar tolerance and median filtering are set to 0 seconds and 11 frames for the LibriMix and SparseLibriMix datasets, and 0.25 seconds and 11 frames for the CALLHOME datasets.
We report SI-SDR improvement (SI-SDRi (dB)), Power (dB/s), SDR improvement (SDRi (dB)), Short-Time Objective Intelligibility (STOI) \cite{stoi} and Perceptual Evaluation of Speech Quality (PESQ) \cite{pesq} for speaker extraction performance.
In terms of complexity analysis, we employ Multiply-Accumulate Operations (MACs (G)) to assess the computational cost of the model and perform a calculation of the model's parameter count.

\begin{table*}[!ht]
\begin{center}
\caption{\label{exp_librimix_max} Performance on the LibriMix dataset for \textbf{max mode}. The results of baseline systems are obtained by our own implementations.}

\begin{tabular}{@{\extracolsep{3pt}}l|cccc|ccc}
\toprule
\multicolumn{8}{c}{\textit{Speaker Diarization}} \\
\midrule
\midrule
\multirow{2}{*}{Model} & \multicolumn{4}{c|}{Overall Performance} & SS & SQ\&QS & QQ  \\
 & DER$\downarrow$ & MS$\downarrow$  &FA$\downarrow$  & SC$\downarrow$ & DER$\downarrow$& DER$\downarrow$ & Seconds$\downarrow$   \\
\midrule
wav2vec 2.0 \textsc{Base} \cite{wav2vec2020al} & 7.62 & 2.28 & 4.82 & 0.52& - & - & -  \\
HuBERT \textsc{Base} \cite{hsu2021hubert} & 7.56& 2.40 & 4.81 & \textbf{0.35}& - & - & -  \\
TS-VAD  \cite{Medennikov2020tsvad} & 7.28 & 3.61 & 2.78 & 0.89& 5.71 & 11.63 & 0.08 \\
 \midrule
USED-F &  \textbf{4.75} & \textbf{2.18}& 2.16 & 0.42& \textbf{3.21} & \textbf{9.41} & 0.05 \\
USED & 5.20  &2.68 & \textbf{2.08} & 0.43 & 3.56 & 10.36  & \textbf{0.04} \\
\toprule
\multicolumn{8}{c}{\textit{Speaker Extraction}} \\
\midrule
\midrule
\multirow{2}{*}{Model} & \multicolumn{4}{c|}{Overall Performance} & SS & SQ & QS\&QQ  \\
 & SI-SDRi$\uparrow$ & SDRi$\uparrow$  &STOI$\uparrow$ &PESQ$\uparrow$  & SI-SDRi$\uparrow$& SI-SDRi$\uparrow$& Power$\downarrow$   \\
 \midrule
 SpEx+ \cite{ge20_interspeech} & 9.06 & 10.11 & 0.776 & 1.39 & 8.32 & 5.17 & 55.60\\
USED-F &  \textbf{12.70} & \textbf{13.22} & \textbf{0.844} & \textbf{1.46} & \textbf{12.00} & 9.17 & \textbf{-24.00}  \\
USED & 12.22  & 12.69  &0.836  & 1.43 & 11.63 & \textbf{9.30} & -14.15 \\
\bottomrule
\end{tabular}
\end{center}

\end{table*}

\section{Experimental Results and Analysis}
\label{sec:exp_result}
\subsection{Results on the LibriMix Dataset}
Table\modify{s} \ref{exp_librimix_max} and \ref{exp_librimix_min} show the results on the LibriMix dataset with max mode and min mode, respectively.
As indicated before, min mode mainly contains highly overlapped speech, while a larger proportion of max mode is non-overlapped speech compared to min mode.
We provide detailed results under different scenarios for the max mode as shown in Table \ref{exp_librimix_max}.
Please note that we use the average duration length that anyone is speaking in seconds as the metric of speaker diarization for the \textit{QQ} scenario.

Our analysis involves comparisons with other popular systems from the literature, including TS-VAD \cite{Medennikov2020tsvad}, EEND-EDA \cite{horiguchi2020end}, HuBERT \textsc{Base} \cite{hsu2021hubert} and wav2vec 2.0 \textsc{Base} \cite{wav2vec2020al} for speaker diarization, SpEx+ \cite{ge20_interspeech}, and Conv-TasNet \cite{luo2019tasnet} for speaker extraction and speech separation, and EEND-SS \cite{ueda2022eend} for joint optimization. 
For HuBERT \textsc{Base} and wav2vec 2.0 \textsc{Base}, we firstly follow the SUPERB diarization task \cite{yang21c_interspeech} to train two downstream models for two speakers and three speakers, respectively.
Subsequently, we combine the output results of these two models for evaluation.

The proposed USED model demonstrates superior or comparable performance compared to the baseline systems, as indicated by their respective evaluation metrics in both the min and max modes.
For speaker diarization, our models exhibit remarkable improvements, with a relative reduction in DER of at least 13\% when compared to the prior systems.
For speaker extraction, our models achieve a much lower value in terms of power compared to other baseline systems under QQ and QS scenarios in the max mode.
This phenomenon implies that speaker diarization and SAD loss effectively mitigate background noise during silent intervals, resulting in enhanced performance.

\begin{table}[!ht]
\begin{center}
\caption{\label{exp_librimix_min} Performance on the LibriMix dataset for \textbf{min mode}. $\dag$ indicates the results are obtained by our own implementations.}

\resizebox{\linewidth}{!}{
\begin{tabular}{l|ccccc}
\toprule
\multicolumn{5}{c}{\textit{Speaker Diarization}} \\
\midrule
\midrule
Model & DER  $\downarrow$ & MS  $\downarrow$ & FA $\downarrow$ & SC $\downarrow$ \\
\midrule
EEND-EDA \cite{ueda2022eend} &  10.16 & - & - & - \\
TS-VAD \dag \cite{Medennikov2020tsvad}  &5.30 & 2.32 & 2.83 & 0.15 \\
\midrule
EEND-SS \cite{ueda2022eend} &  6.27 & - & - & - \\
\quad + LMF &  6.04 & - & - & - \\
\midrule
USED-F & 4.64 & 1.99 & 2.55  & 0.10 \\
USED & \textbf{4.57} & \textbf{1.94} & \textbf{2.54} & \textbf{0.09} \\

\toprule
\multicolumn{5}{c}{\textit{Speaker Extraction / Speech Separation}} \\
\midrule
\midrule
Model & SI-SDRi  $\uparrow$ & SDRi $\uparrow$ & STOI $\uparrow$ & PESQ  $\uparrow$  \\
\midrule
Conv-TasNet \cite{ueda2022eend} & 7.66 & 8.71 & 0.756 & - \\
SpEx+ \dag \cite{ge20_interspeech} & 8.76 & 10.09 & 0.772 & 1.44 \\
\midrule
EEND-SS \cite{ueda2022eend} & 9.31 & 7.50 & 0.760 & - \\
\quad + Fusion & 9.38 & 7.59 & 0.760 & - \\
\quad + LMF & 8.83 & 9.72 & 0.767 & - \\
\quad + LMF + Fusion & 8.87 & 9.77 & 0.767 & - \\
\midrule
USED-F & \textbf{12.24} & \textbf{12.83} & \textbf{0.841} & \textbf{1.54} \\
USED & 11.98 & 12.50 & 0.834 & 1.44 \\
\bottomrule
\end{tabular}
}
\end{center}

\end{table}

\subsection{Results on SparseLibriMix}
\begin{figure}%
    \centering
    \subfloat[\centering Speaker Diarization]{{\includegraphics[width=0.485\linewidth]{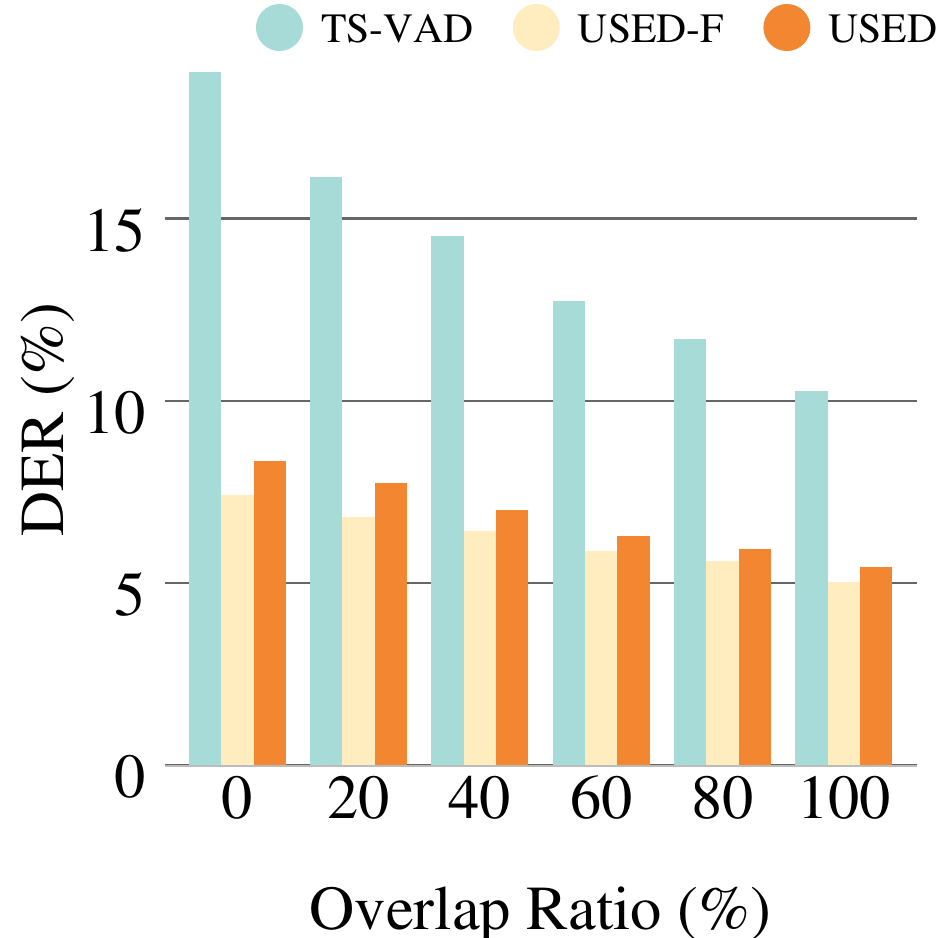} }} \hfill
    \subfloat[\centering Speaker Extraction]{{\includegraphics[width=0.485\linewidth]{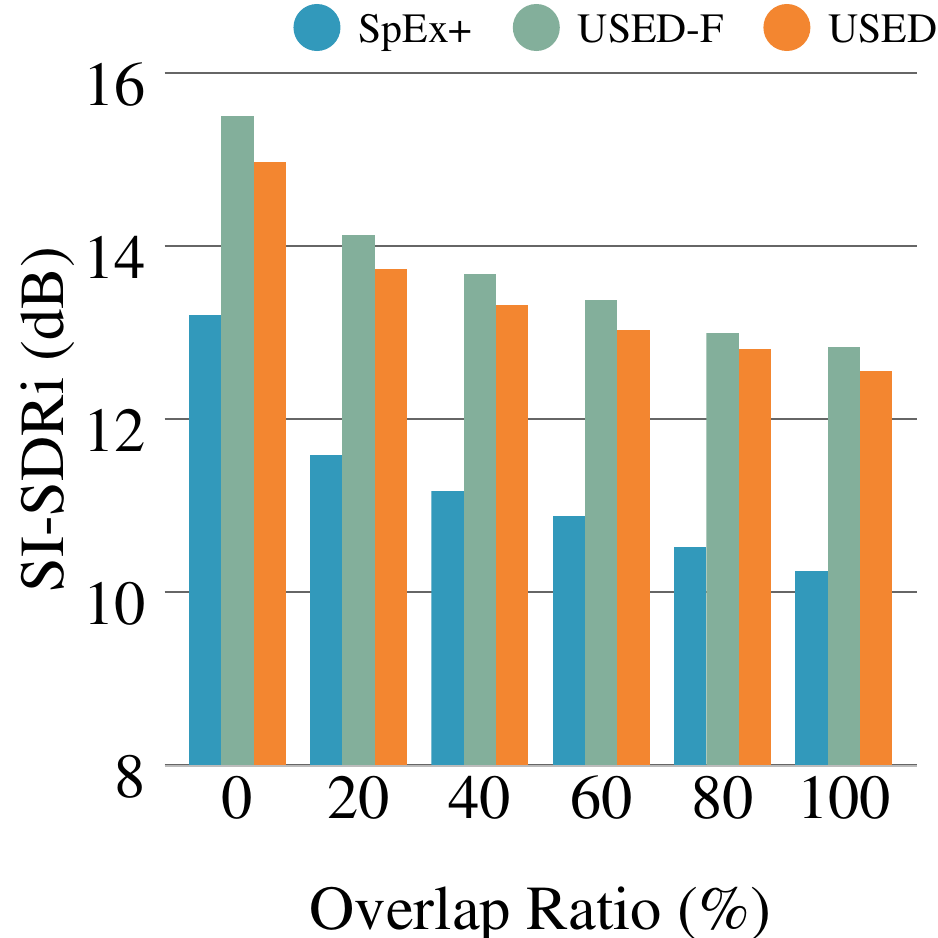} }}%
    \caption{Performance on the SparseLibriMix for overlap ratio from 0\% to 100\%. The Left and right bar charts are the results of speaker diarization and speaker extraction tasks, respectively. }%
    \label{fig:sparse}%
\vspace{-10pt}
\end{figure}

In addition to the LibriMix dataset, we are also interested in assessing the performance of our model across a range of overlap ratios.
Therefore, we conduct an evaluation on SparseLibriMix, as illustrated in Fig. \ref{fig:sparse}.
We select TS-VAD and SpEx+ models as the baseline systems.
The model trained on the max mode of Libri2Mix and Libri3Mix is evaluated on SparseLibriMix with 2 and 3 speakers.
The proposed USED models outperform each \modify{task baseline},  confirming our models' effectiveness for variable values of overlap ratio.

\subsection{Results on CALLHOME}
To compare the performance of diarization on real data, we evaluate our model on the CALLHOME dataset.
The results are summarized in Table \ref{callhome}, compared with several speaker diarization methods, including two clustering-based methods, AHC clustering \cite{horiguchi2020end} and spectral clustering \cite{8462628}, PixIT \cite{kalda2024pixit}, end-to-end speaker diarization models, EEND-EDA \cite{horiguchi2020end}, SC-EEND \cite{fujita2020neural} and AED-EEND \cite{chen23n_interspeech}, and the TS-VAD model \cite{Medennikov2020tsvad}.
Most of those methods use the simulation data from Swichboard-2, Switchboard Cellular, and NIST Speaker Recognition Evaluation \cite{horiguchi2020end}, which have a variable number of speakers, ranging from one to five.
The total number of hours for the pre-training data is around 15,516 hours for those methods.
In contrast, we use much less simulation data, which is from LibriMix and only have 2 or 3 speakers.
This results in around 452 hours of pre-training data.

Our USED model achieves around 15\% relative improvement on overall DER compared to the TS-VAD baseline and performs better with much less pre-training data compared to other methods.
We observe that our USED model is more robust in terms of the number of speakers.
Results show that although our model only uses simulation data from LibriMix, which has a speech mixture with 2 or 3 speakers, the model still achieves better DER results when the number of speakers is larger than 3.
We believe this gain is from the design of the embedding assignment module.

\begin{table*}[!ht]
\begin{center}
\caption{\label{callhome} DER (\%) results on CALLHOME, a dataset of real recordings. The symbol $\ddagger$ denotes results obtained from our own implementations, while $\dag$ indicates results evaluated using the open-source model.}

\begin{tabular}{@{\extracolsep{3pt}}l|c|ccccccc}
\toprule
\multirow{2}{*}{Model} & \multirow{2}{*}{Pre-training Data} & \multicolumn{6}{c}{Number of Speakers} \\
 &  & 2 & 3 & 4 & 5 & 6 & all \\
\midrule
\midrule
AHC clustering \cite{horiguchi2020end} & - & 15.45 & 18.01 & 22.68 & 31.40 & 34.27 & 19.43 \\
spectral clustering \cite{8462628} & - & 16.05 & 18.77 & 22.05 & 30.46 & 36.85 & 19.83 \\
PixIT  \dag \cite{kalda2024pixit} & 79h & 17.02 & 26.03 & 27.75 & 36.38 & 40.77 & 24.41 \\
EEND-EDA \cite{horiguchi2020end} & 15,516h & 8.50 & 13.24 & 21.46 & 33.16 & 40.29 & 15.29 \\
SC-EEND \cite{fujita2020neural} & 15,516h & 9.57 & 14.00 & 21.14 & 31.07 & 37.06 & 15.75 \\
AED-EEND \cite{chen23n_interspeech} & 15,516h & \textbf{6.96} & \textbf{12.56} & 18.26 & 34.32 & 44.52 & 14.22 \\
TS-VAD $\ddagger$ \cite{Medennikov2020tsvad} & 452h & 12.95 & 14.57 & 20.26 & 27.36 & 32.66 & 15.51 \\
\midrule
USED-F & 452h & 9.09 & 12.76 & \textbf{14.67} & \textbf{26.96} & 26.71 & \textbf{13.16}  \\
USED & 452h & 10.23 & 12.78 & 14.68 & 32.02 & \modify{\textbf{25.10}} & 13.51 \\
\bottomrule
\end{tabular}
\end{center}
\vspace{-15pt}
\end{table*}

\begin{table}[!ht]
\begin{center}
\caption{\label{exp_control} Model performance on the LbriMix dataset for max mode when setting different values of $m$. RS stands for residual state.}
\begin{tabular}{c|l|cc}
\toprule
$m$ & Model&  SI-SDRi $\uparrow$ & DER $\downarrow$ \\
\midrule
\midrule
\multirow{5}{*}{1} &SpEx+ \cite{ge20_interspeech}   & 9.06 & - \\
&USED-F  &5.71 & 20.30 \\
& USED& 9.73 & 11.13 \\
& \quad - w/o RS & 9.87 &  11.96 \\
& \quad - $p_{\text{res}}=1.0$  & \textbf{10.23} & \textbf{10.85}  \\
\midrule
\multirow{4}{*}{2} & USED-F  & 9.55 & 11.27 \\
& USED   & 11.02 & \textbf{9.42} \\
& \quad - w/o RS   & 10.96  &  10.98 \\
& \quad - $p_{\text{res}}=1.0$  & \textbf{11.08} & 10.14 \\ 
\midrule
\multirow{5}{*}{all} &TS-VAD \cite{Medennikov2020tsvad}   & - & 7.28 \\
&USED-F  & \textbf{12.70} & \textbf{4.75} \\
& USED  & 12.22 & 5.20 \\
& \quad - w/o RS  & 11.80 &  9.46 \\
& \quad - $p_{\text{res}}=1.0$ &  11.76  & 10.38  \\
\bottomrule
\end{tabular}
\end{center}
\vspace{-15pt}
\end{table}

\subsection{Investigation of USED Models for Generating Results of Different Numbers of Speakers During Inference}

In this section, we compare the USED-F model and the USED model with baseline systems when generating results for different numbers of speakers during inference, as shown in Table \ref{exp_control}.
As introduced in Section \ref{sec:loss_se}, $m$ is the number of embeddings with an \textit{active} state as the input of the USED model. During inference, $m$ also represents the number of speakers the USED model is required to predict.
When $m$ is 1, USED models perform speaker extraction like SpEx+.
Under this condition, USED models and SpEx+ have the same input data for a fair comparison. 
When $m$ is all, it means we provide all speech references for present speakers, and the embedding is $V_{\text{em}}$ for \textit{residual} state.
Otherwise, we use $V_{\text{res}}$ as the embedding for \textit{residual} state.

First, USED models consistently perform much better than the SpEx+ baseline when $m$ is 1, except for the USED-F model.
This indicates that even if the USED model only gets one speech reference, it can still outperform the SpEx+ model.
When $m$ is 2 or all, the performance of USED in speaker extraction significantly surpasses that of SpEx+.
\modify{The experimental results show that the USED model demonstrates superior efficiency and performance in speaker extraction tasks, particularly in multi-speaker scenarios.
Unlike SpEx+, which requires multiple inferences for each speaker, USED performs inference just once, reducing redundant computations like those in the speech encoder.
It can also simultaneously utilize the embeddings of multiple speakers, resulting in more discriminative results, whereas SpEx+ processes one speaker at a time.
As the number of speakers increases, USED’s performance advantage becomes even more significant.}

\modify{Second, when extracting results for all speakers, the USED model outperforms TS-VAD, as shown in the experimental results.
In addition, the TS-VAD model can only extract results for all speakers simultaneously, while the USED model provides flexibility to extract results for a specific speaker or multiple speakers, which is advantageous in scenarios where obtaining embeddings for all speakers at once is not feasible.
Furthermore, the USED model does not require a pre-trained speaker encoder, unlike TS-VAD, which typically relies on one.
}

Third, we observe that the USED model without \textit{residual} state achieves improvement compared to the USED-F model on both speaker diarization and extraction tasks if $m$ equals 1 or 2.
It gets a larger improvement when applying \textit{residual} state with $p_{\text{res}}$ of 1.0.
However, when $m$ is all, the performance of these two models degrades significantly on the speaker diarization task compared to the USED-F model \modify{with} the DER \modify{increasing} from 4.75\% to 10.38\% and 9.46\%, \modify{respectively}.
This degradation can be avoided by setting $p_{\text{res}}$ to 0.9 during training.
In this way, the USED model can achieve a \modify{competitive} performance compared to \modify{that of} the USED-F model if $m$ is all and still significantly outperforms baseline systems when $m$ is 1 or 2.

The performance gap between USED-F and USED models when $m = \text{all}$ is due to the fact that the input to the USED-F model consistently includes speaker embeddings for all speakers during training.
As a result, USED-F is designed to predict for all speakers simultaneously, making it specifically optimized for scenarios where $m = \text{all}$.
However, its performance is less robust in cases where $m \neq \text{all}$ (e.g., $m=1$ or $m=2$), leading to lower performance compared to the USED model in these situations.

\begin{table*}[!ht]
\begin{center}
\caption{\label{exp_int_sad} USED-F model performance under different scenarios when using different loss weights for each scenario of SAD loss with or without Multi-Task interaction module. IM means multi-task interaction module. The results of the first line are from the standard setting.
}
\begin{tabular}{c|c|c|c|c|cccc|c}
\toprule
 \multirow{3}{*}{$\alpha$} &  \multirow{3}{*}{$\beta$} & \multirow{3}{*}{$\gamma$}  & \multirow{3}{*}{$\delta$} & \multirow{3}{*}{IM}  & \multicolumn{4}{c|}{\textbf{Speaker Extraction}} &\textbf{Speaker Diarization}  \\
  &  & &   & &  SI-SDRi $\uparrow$  &  SI-SDRi $\uparrow$&  SI-SDRi $\uparrow$& Power $\downarrow$ & \multirow{2}{*}{DER$\downarrow$}\\
 &  & &   &  &  \textit{All} &  \textit{SS} & \textit{SQ}              & \textit{QS \& QQ}  &  \\
\midrule
\midrule
\modify{0.001} & \modify{0.001} & \modify{1.0} & \modify{1.0} &  \cmark & 12.70 & 12.00 & 9.17 & -24.00 & 4.75 \\
\midrule
\modify{0} & \modify{0} & \modify{1.0} & \modify{1.0}  &  \xmark & 12.47 & 12.02 & 9.58 & 55.68 & 4.80 \\
\modify{0.001} & \modify{0.001} & \modify{1.0} & \modify{1.0}  & \xmark & 12.71 & 12.00 & 9.11 & -1.93 & 4.65 \\
\modify{0.01} & \modify{0.01} & \modify{1.0} & \modify{1.0}  & \xmark & 12.50 & 11.65 & 8.68 & -45.16 & 4.77 \\
\modify{1.0} & \modify{1.0} & \modify{1.0} & \modify{1.0}  &  \xmark & 11.54 & 10.48 & 7.42 & -79.21 & 5.57 \\
\bottomrule
\end{tabular}
\end{center}
\vspace{-12pt}
\end{table*}

\begin{table}[!ht]
\begin{center}
\caption{\label{exp_single} Ablation study of the USED model configured for a single task (Only SD or
Only SE) on the LibriMix dataset for max mode.}

\begin{tabular}{@{\extracolsep{3pt}}l|cccc}
\toprule
\multicolumn{5}{c}{\textit{Speaker Diarization}} \\
\midrule
\midrule
\multirow{2}{*}{Model} & All & SS & SQ\&QS & QQ  \\
 & DER$\downarrow$ &DER$\downarrow$& DER$\downarrow$ & Seconds$\downarrow$   \\
\midrule
USED-F &  4.75 & 3.21 & 9.41 & 0.05 \\
\quad - Only SD & 6.43 & 4.72 & 11.76 & 0.04 \\
\toprule
\multicolumn{5}{c}{\textit{Speaker Extraction}} \\
\midrule
\midrule
\multirow{2}{*}{Model} & All & SS & SQ & QS\&QQ  \\
 & SI-SDRi$\uparrow$ & SI-SDRi$\uparrow$& SI-SDRi$\uparrow$& Power$\downarrow$   \\
 \midrule
USED-F &  12.70 &  12.00 & 9.17 & -24.00 \\
\quad - Only SE & 12.46 & 11.71 & 9.03 & 20.84 \\
\bottomrule
\end{tabular}
\end{center}
\vspace{-15pt}
\end{table}

\subsection{Ablation Study: Assessing Single-Task Training for the USED Model}
We then compare USED models trained on single task to investigate the impact of each task as illustrated in Table \ref{exp_single}.
Without loss of generality, we use the USED-F model for comparison.
We present the results obtained by setting $\lambda_1$ or $\lambda_2$ to 0, referred to as Only SD (Speaker Diarization) and Only SE (Speaker Extraction), respectively.

The USED-F model achieves a notable reduction of DER with a relative improvement of around 32\% in the SS scenario and 20\% in the SQ\&QS scenarios. 
The larger improvement in the SS scenario suggests that speaker extraction indeed plays a pivotal role in enhancing the diarization module's ability to handle overlapping speech segments.
Furthermore, SI-SDRi increases from 12.46 dB to 12.70 dB for overall performance when compared to the USED-F model with only SE, which may be mainly from the improvement under the QQ scenario where the power reduces from 20.84 dB/s to -24.00 dB/s.
This indicates that the speaker diarization task helps surpass the background noise for the speaker extraction task.

\subsection{Investigation of the Relationship between the Mult\modify{i}-Task Interaction Module and the SAD Loss}

\begin{table}[!ht]
\begin{center}
\caption{Model Comparison in Terms of Parameters and MACs for the TS-VAD model, the SpEx+ model and the USED model. ``Params'' refers to the \modify{number of} model parameters, which is counted in millions (M). \dag The MACs of SpEx+ is only for extracting one speaker's result. \label{tab:time_com}}
\begin{tabular}{c|ccc}
\toprule
Model & Params & MACs & Total MACs \\
\midrule
\midrule
TS-VAD \cite{Medennikov2020tsvad} & 39.50 & 27.21  & \multirow{2}{*}{158.01} \\
SpEx+ \dag \cite{ge20_interspeech} & 16.35  & 43.60 \\
\midrule
USED-F& 23.12 & 96.91  & 96.91 \\
\midrule
USED& 23.65  & 119.54 & 119.54 \\
\bottomrule
\end{tabular}
\end{center}
\vspace{-12pt}
\end{table}

In this section, we analyze the performance of the USED-F model using different weights for each scenario of SAD loss with or without the multi-task interaction module as illustrated in Table \ref{exp_int_sad}.
For the experiments without the multi-task interaction module, results demonstrate that when increasing the weights \modify{$\alpha$ and $\beta$ for \textit{QQ} and \textit{QS}} scenarios, we can get a lower power value, but the SI-SDRi are reduced from 12.02 dB to 10.48 dB for \modify{the} \textit{SS} scenario and from 9.58 dB to 7.42 dB for \modify{the} \textit{SQ} scenario.
Regarding the performance of speaker diarization, DER results increase when larger values for the weights \modify{$\alpha$ and $\beta$ of \textit{QQ} and \textit{QS}} scenarios are used.
These observations indicate that the model performance is sensitive to the loss weights under different scenarios.
With the help of the multi-task interaction module, we can keep the performance for speaker extraction under \textit{SS} and \textit{SQ} scenarios and speaker diarization \modify{and simultaneously} achieve a much lower power \modify{under the QQ scenario}, which decreases from -1.93 dB/s to -24.00 dB/s. 

\subsection{Complexity Analysis}
In this section, we conduct a complexity analysis of the USED model in comparison to the baseline models, TS-VAD and SpEx+, as shown in Table \ref{tab:time_com}.
To ensure a fair comparison, the duration of the speech input for all models is kept consistent, with both the speech mixture and the speech references set to 4 seconds.
The experimental results indicate that, compared to individual baseline models, the USED model has more parameters than the SpEx+ model but fewer than the TS-VAD model.
However, the computational complexity of the USED model is higher than both SpEx+ and TS-VAD.

\begin{table*}[!ht]
\begin{center}
\caption{Impact of hyperparameter choices for embedding assignment module on model performance.\label{tab:hyp}}
\begin{tabular}{cc|ccccccccccccccc}
\toprule
\multirow{2}{*}{$m$} & \multirow{2}{*}{Metrics} & \multicolumn{5}{c}{$p_{sel}$} & \multicolumn{5}{c}{$p_{em}$}  & \multicolumn{5}{c}{$p_{res}$} \\
\cmidrule(l){3-7} \cmidrule(l){8-12} \cmidrule(l){13-17}
 & & 0.1 & 0.3 & 0.5 & 0.7 & 0.9 & 0.1 & 0.3 & 0.5 & 0.7 & 0.9 & 0.1 & 0.3 & 0.5 & 0.7 & 0.9  \\
\midrule
\midrule
\multirow{2}{*}{1} & SI-SDRi $\uparrow$  & 11.81 & 10.36 & 12.12& 12.81 & 12.60 & 10.77  &
12.12 & 12.93 & 13.09 & 13.10 & 13.12 & 12.99 & 12.72 & 13.01 &12.12 \\
& DER $\downarrow$ & 6.80 & 8.79 & 6.47 & 5.83 & 6.19 & 6.67 & 6.47 & 5.87 & 5.63 & 5.62 & 5.75 & 5.83 & 6.21 & 5.76 & 6.47 \\
\midrule
\multirow{2}{*}{2} & SI-SDRi $\uparrow$  & 13.28 & 13.30 & 13.58 & 13.62 & 12.81 & 13.68 & 13.58 & 13.83 & 13.56 & 13.42 & 13.77 & 13.76 & 13.83 & 13.72 & 13.58 \\
& DER $\downarrow$ & 5.17 & 5.11 & 4.98& 5.28 & 9.41 & 5.01 &4.98 & 5.32 & 4.97 & 5.05 & 5.67 & 5.15 & 5.24 & 4.93 & 4.98\\
\midrule
\multirow{2}{*}{all} & SI-SDRi $\uparrow$  & 14.71 & 14.45 &14.29 & 14.07 & 13.10 & 14.25 &14.29& 14.31 & 14.26 & 14.27 & 14.31 & 14.28 & 14.35 & 14.25 & 14.29 \\
& DER $\downarrow$ & 3.87 & 4.00 &4.22 & 4.41 & 6.27 & 4.30 &4.22& 4.16  & 4.13 & 4.18 & 4.23 & 4.22 & 4.16 & 4.19 &4.22\\
\bottomrule
\end{tabular}
\end{center}
\end{table*}

\begin{table}[!ht]
\begin{center}
\caption{Model performance regarding different loss weight values on the validation set. \label{tab:loss_weight}}
\resizebox{\linewidth}{!}{
\begin{tabular}{cc|ccccccccc}
\toprule
\multirow{2}{*}{$\lambda_1$} &\multirow{2}{*}{$\lambda_2$}  & \multicolumn{2}{c}{$\lambda_3 = 0.1$} & \multicolumn{2}{c}{$\lambda_3 = 0.5$} & \multicolumn{2}{c}{$\lambda_3 = 1.0$} \\
\cmidrule(l){3-4} \cmidrule(l){5-6} \cmidrule(l){7-8}
 &  & SI-SDRi $\uparrow$ & DER  $\downarrow$  & SI-SDRi $\uparrow$ & DER  $\downarrow$  & SI-SDRi $\uparrow$ & DER  $\downarrow$  \\
 \midrule
0.1 & 0.1 & 14.31 & 4.15 & 14.27 & 4.26 & 14.27 & 4.15 \\
0.1 & 0.5 & 14.01 & 3.85 & 13.99 & 3.76 & 14.06 & 3.74 \\
0.1 & 1.0 & 13.54 & 3.85 & 13.46 & 4.00 & 13.32 & 4.19 \\
0.5 & 0.1 & 14.22 & 5.01 & 14.13 & 5.14 & 14.19 & 5.01  \\
0.5 & 0.5 & 14.16 & 4.23 & 14.27 & 4.26 & 14.32 & 4.14 \\
0.5 & 1.0 & 14.19 & 3.97 & 14.32 & 3.92 & 14.28 & 3.84 \\
1.0 & 0.1 & 14.11 & 5.49 & 14.15 & 5.37 & 14.17 & 5.47 \\
1.0 & 0.5 & 14.15 & 4.57 & 14.18 & 4.55 & 14.32 & 4.45 \\
1.0 & 1.0 & 14.22 & 4.21 & 14.32 & 4.14 & 14.29 & 4.22 \\

\bottomrule
\end{tabular}
}
\end{center}
\vspace{-12pt}
\end{table}

It is important to note that the USED model simultaneously addresses two tasks: speaker extraction and diarization, whereas TS-VAD and SpEx+ are designed for a single task, with SpEx+ specifically extracting the speech for only one speaker.
Therefore, we calculate the `Total MACs'.
`Total MACs' refers to the total number of MACs required for the models to process a speech mixture containing three speakers and obtain the corresponding speaker extraction and diarization results for all three speakers.
When comparing the `Total MACs', the `Total MACs' of the USED and USED-F models are significantly lower than that of the baseline models \modify{combined}. Specifically, the USED-F model reduces the total MACs by approximately 39\%, while the USED model achieves a reduction of around 24\%.

\subsection{Loss Weight Selection}
To investigate the impact of different loss weights on the model's performance and explore how to select appropriate values for these weights, we conduct a series of experiments with three loss weights ($\lambda_1$, $\lambda_2$, and $\lambda_3$), as illustrated in Table \ref{tab:loss_weight}.
We evaluate the model's performance on the validation set of the LibriMix dataset with max mode by testing each loss weight with values of 0.1, 0.5, and 1.0.
The experimental results indicate that there may be room for improvement in the model's performance regarding the selection of loss weights, although we set the values of all three weights to 1.0 for other experiments.
\modify{For example, when $\lambda_1$, $\lambda_2$, and $\lambda_3$ are 0.5, 1.0 and 0.5, respectively}, the USED model performs slightly better on both the speaker extraction and diarization tasks.

\subsection{Impact \modify{on Model Performance} of Hyperparameter Tuning for Embedding Assignment Module}
In this section, we examine the influence of hyperparameter selection within the embedding assignment module on the overall performance of the model, as shown in Table \ref{tab:hyp}.
Specifically, we focus on three key hyperparameters associated with the embedding assignment module, conducting our analysis using the validation set from the LibriMix dataset in max mode.
We assess the impact of varying the hyperparameter values $p_{sel}$, $p_{em}$, and $p_{res}$, which are assigned values of 0.1, 0.3, 0.5, 0.7, and 0.9, \modify{while keeping the two other hyperparameters fixed to their default value}, across multiple configurations where $m$ is set to 1, 2, and all.

The experimental results demonstrate that the influence of the three hyperparameters on model performance varies depending on the value of $m$.
For instance, for hyperparameter $p_{sel}$, larger values lead to better results when $m$ equals 1, whereas smaller values yield better performance when $m$ equals all.
After carefully evaluating the overall performance across different values of $m$, we selected a set of parameters with $p_{sel} = 0.5$, $p_{em} = 0.3$, and $p_{res} = 0.9$ for relatively good performance.

\section{Conclusion}
\label{sec:conclu}
We propose USED, a unified network for universal speaker extraction and diarization, which integrates speaker extraction and diarization for managing speech mixture\modify{s} with varying overlap ratios \modify{and variable number of speakers}.
We design an embedding assignment module to support producing results for a variable number of speakers based on speech references and enhance the model's robustness.
Additionally, a multi-task interaction module is designed to leverage information from both speaker extraction and diarization tasks.
Experimental results demonstrate significant improvements in both highly and sparsely overlapped speech scenarios for the speaker extraction and speaker diarization tasks.

\bibliographystyle{IEEEtran}
\bibliography{refs}

\begin{thebibliography}{10}
\providecommand{\url}[1]{#1}
\csname url@samestyle\endcsname
\providecommand{\newblock}{\relax}
\providecommand{\bibinfo}[2]{#2}
\providecommand{\BIBentrySTDinterwordspacing}{\spaceskip=0pt\relax}
\providecommand{\BIBentryALTinterwordstretchfactor}{4}
\providecommand{\BIBentryALTinterwordspacing}{\spaceskip=\fontdimen2\font plus
\BIBentryALTinterwordstretchfactor\fontdimen3\font minus \fontdimen4\font\relax}
\providecommand{\BIBforeignlanguage}[2]{{%
\expandafter\ifx\csname l@#1\endcsname\relax
\typeout{** WARNING: IEEEtran.bst: No hyphenation pattern has been}%
\typeout{** loaded for the language `#1'. Using the pattern for}%
\typeout{** the default language instead.}%
\else
\language=\csname l@#1\endcsname
\fi
#2}}
\providecommand{\BIBdecl}{\relax}
\BIBdecl

\bibitem{zmolikova2023neural}
K.~Zmolikova, M.~Delcroix, T.~Ochiai, K.~Kinoshita, J.~{\v{C}}ernock{\`y}, and D.~Yu, ``Neural target speech extraction: An overview,'' \emph{IEEE Signal Processing Magazine}, vol.~40, no.~3, pp. 8--29, 2023.

\bibitem{park2022review}
T.~J. Park, N.~Kanda, D.~Dimitriadis, K.~J. Han, S.~Watanabe, and S.~Narayanan, ``A review of speaker diarization: Recent advances with deep learning,'' \emph{Computer Speech \& Language}, vol.~72, p. 101317, 2022.

\bibitem{rao19_interspeech}
W.~Rao, C.~Xu, E.~S. Chng, and H.~Li, ``Target speaker extraction for multi-talker speaker verification,'' in \emph{Proc. Interspeech 2019}, 2019, pp. 1273--1277.

\bibitem{xu2021sv}
C.~Xu, W.~Rao, J.~Wu, and H.~Li, ``Target speaker verification with selective auditory attention for single and multi-talker speech,'' \emph{IEEE/ACM Transactions on Audio, Speech, and Language Processing}, vol.~29, pp. 2696--2709, 2021.

\bibitem{zmolikova2017speaker}
K.~Žmolíková, M.~Delcroix, K.~Kinoshita, T.~Higuchi, A.~Ogawa, and T.~Nakatani, ``Speaker-aware neural network based beamformer for speaker extraction in speech mixtures,'' in \emph{Proc. Interspeech 2017}, 2017, pp. 2655--2659.

\bibitem{marc2018seforasr}
M.~Delcroix, K.~Zmolikova, K.~Kinoshita, A.~Ogawa, and T.~Nakatani, ``Single channel target speaker extraction and recognition with speaker beam,'' in \emph{2018 IEEE International Conference on Acoustics, Speech and Signal Processing (ICASSP)}, 2018, pp. 5554--5558.

\bibitem{radford2023robust}
A.~Radford, J.~W. Kim, T.~Xu, G.~Brockman, C.~McLeavey, and I.~Sutskever, ``Robust speech recognition via large-scale weak supervision,'' in \emph{International conference on machine learning}.\hskip 1em plus 0.5em minus 0.4em\relax PMLR, 2023, pp. 28\,492--28\,518.

\bibitem{mansfield21_interspeech}
C.~Mansfield, S.~Ng, G.-A. Levow, R.~A. Wright, and M.~Ostendorf, ``Revisiting parity of human vs. machine conversational speech transcription,'' in \emph{Interspeech 2021}, 2021, pp. 1997--2001.

\bibitem{ge20_interspeech}
M.~Ge, C.~Xu, L.~Wang, E.~S. Chng, J.~Dang, and H.~Li, ``{S}p{E}x+: A complete time domain speaker extraction network,'' in \emph{Proc. Interspeech 2020}, 2020, pp. 1406--1410.

\bibitem{xu2020spex}
C.~Xu, W.~Rao, E.~S. Chng, and H.~Li, ``{S}p{E}x: Multi-scale time domain speaker extraction network,'' \emph{IEEE/ACM Transactions on Audio, Speech, and Language Processing}, vol.~28, pp. 1370--1384, 2020.

\bibitem{liu2023x}
K.~Liu, Z.~Du, X.~Wan, and H.~Zhou, ``{X}-{S}ep{F}ormer: End-to-end speaker extraction network with explicit optimization on speaker confusion,'' in \emph{ICASSP 2023 - 2023 IEEE International Conference on Acoustics, Speech and Signal Processing (ICASSP)}.\hskip 1em plus 0.5em minus 0.4em\relax IEEE, 2023, pp. 1--5.

\bibitem{Fujita2019eend}
Y.~Fujita, N.~Kanda, S.~Horiguchi, K.~Nagamatsu, and S.~Watanabe, ``End-to-end neural speaker diarization with permutation-free objectives,'' in \emph{Proc. Interspeech 2019}, 2019, pp. 4300--4304.

\bibitem{horiguchi2020end}
S.~Horiguchi, Y.~Fujita, S.~Watanabe, Y.~Xue, and K.~Nagamatsu, ``End-to-end speaker diarization for an unknown number of speakers with encoder-decoder based attractors,'' in \emph{Proc. Interspeech 2020}, 2020, pp. 269--273.

\bibitem{Medennikov2020tsvad}
I.~Medennikov, M.~Korenevsky, T.~Prisyach, Y.~Khokhlov, M.~Korenevskaya, I.~Sorokin, T.~Timofeeva, A.~Mitrofanov, A.~Andrusenko, I.~Podluzhny, A.~Laptev, and A.~Romanenko, ``Target-speaker voice activity detection: A novel approach for multi-speaker diarization in a dinner party scenario,'' in \emph{Proc. Interspeech 2020}, 2020, pp. 274--278.

\bibitem{jiang2023prompt}
Y.~Jiang, Z.~Chen, R.~Tao, L.~Deng, Y.~Qian, and H.~Li, ``Prompt-driven target speech diarization,'' in \emph{ICASSP 2024 - 2024 IEEE International Conference on Acoustics, Speech and Signal Processing (ICASSP)}, 2024, pp. 11\,086--11\,090.

\bibitem{chen2023attention}
Z.~Chen, B.~Han, S.~Wang, and Y.~Qian, ``Attention-based encoder-decoder end-to-end neural diarization with embedding enhancer,'' \emph{IEEE/ACM Transactions on Audio, Speech, and Language Processing}, vol.~32, pp. 1636--1649, 2024.

\bibitem{chen2020continuous}
Z.~Chen, T.~Yoshioka, L.~Lu, T.~Zhou, Z.~Meng, Y.~Luo, J.~Wu, X.~Xiao, and J.~Li, ``Continuous speech separation: Dataset and analysis,'' in \emph{ICASSP 2020-2020 IEEE International Conference on Acoustics, Speech and Signal Processing (ICASSP)}.\hskip 1em plus 0.5em minus 0.4em\relax IEEE, 2020, pp. 7284--7288.

\bibitem{shafey2019joint}
L.~E. Shafey, H.~Soltau, and I.~Shafran, ``Joint speech recognition and speaker diarization via sequence transduction,'' in \emph{Interspeech 2019}, 2019, pp. 396--400.

\bibitem{kanda2020joint}
N.~Kanda, Y.~Gaur, X.~Wang, Z.~Meng, Z.~Chen, T.~Zhou, and T.~Yoshioka, ``Joint speaker counting, speech recognition, and speaker identification for overlapped speech of any number of speakers,'' in \emph{Interspeech 2020}, 2020, pp. 36--40.

\bibitem{kanda2021minimum}
N.~Kanda, Z.~Meng, L.~Lu, Y.~Gaur, X.~Wang, Z.~Chen, and T.~Yoshioka, ``Minimum {B}ayes risk training for end-to-end speaker-attributed {ASR},'' in \emph{ICASSP 2021-2021 IEEE International Conference on Acoustics, Speech and Signal Processing (ICASSP)}.\hskip 1em plus 0.5em minus 0.4em\relax IEEE, 2021, pp. 6503--6507.

\bibitem{cosentino2020librimix}
J.~Cosentino, M.~Pariente, S.~Cornell, A.~Deleforge, and E.~Vincent, ``{L}ibri{M}ix: An open-source dataset for generalizable speech separation,'' \emph{arXiv preprint arXiv:2005.11262}, 2020.

\bibitem{cetin06_interspeech}
Özgür Çetin and E.~Shriberg, ``Analysis of overlaps in meetings by dialog factors, hot spots, speakers, and collection site: insights for automatic speech recognition,'' in \emph{Proc. Interspeech}, 2006.

\bibitem{barker18_interspeech}
J.~Barker, S.~Watanabe, E.~Vincent, and J.~Trmal, ``The fifth '{CHiME}' speech separation and recognition challenge: Dataset, task and baselines,'' in \emph{Proc. Interspeech 2018}, 2018, pp. 1561--1565.

\bibitem{boeddeker2023ts}
C.~Boeddeker, A.~S. Subramanian, G.~Wichern, R.~Haeb-Umbach, and J.~Le~Roux, ``{TS-SEP}: Joint diarization and separation conditioned on estimated speaker embeddings,'' \emph{IEEE/ACM Transactions on Audio, Speech, and Language Processing}, vol.~32, pp. 1185--1197, 2024.

\bibitem{ueda2022eend}
S.~Maiti, Y.~Ueda, S.~Watanabe, C.~Zhang, M.~Yu, S.-X. Zhang, and Y.~Xu, ``{EEND-SS}: Joint end-to-end neural speaker diarization and speech separation for flexible number of speakers,'' in \emph{2022 IEEE Spoken Language Technology Workshop (SLT)}, 2023, pp. 480--487.

\bibitem{yu2017permutation}
D.~Yu, M.~Kolb{\ae}k, Z.-H. Tan, and J.~Jensen, ``Permutation invariant training of deep models for speaker-independent multi-talker speech separation,'' in \emph{2017 IEEE International Conference on Acoustics, Speech and Signal Processing (ICASSP)}.\hskip 1em plus 0.5em minus 0.4em\relax IEEE, 2017, pp. 241--245.

\bibitem{zexu2022usev}
Z.~Pan, M.~Ge, and H.~Li, ``{USEV}: Universal speaker extraction with visual cue,'' \emph{IEEE/ACM Transactions on Audio, Speech, and Language Processing}, vol.~30, pp. 3032--3045, 2022.

\bibitem{luo2019tasnet}
Y.~Luo and N.~Mesgarani, ``{C}onv-{T}as{N}et: Surpassing ideal time–frequency magnitude masking for speech separation,'' \emph{IEEE/ACM Transactions on Audio, Speech, and Language Processing}, vol.~27, no.~8, pp. 1256--1266, 2019.

\bibitem{yi2020dprnn}
Y.~Luo, Z.~Chen, and T.~Yoshioka, ``{D}ual-{P}ath {RNN}: Efficient long sequence modeling for time-domain single-channel speech separation,'' in \emph{ICASSP 2020 - 2020 IEEE International Conference on Acoustics, Speech and Signal Processing (ICASSP)}, 2020, pp. 46--50.

\bibitem{luo2023music}
Y.~Luo and J.~Yu, ``Music source separation with band-split {RNN},'' \emph{IEEE/ACM Transactions on Audio, Speech, and Language Processing}, vol.~31, pp. 1893--1901, 2023.

\bibitem{zk2017spkbeam1}
K.~Žmolíková, M.~Delcroix, K.~Kinoshita, T.~Higuchi, A.~Ogawa, and T.~Nakatani, ``Learning speaker representation for neural network based multichannel speaker extraction,'' in \emph{2017 IEEE Automatic Speech Recognition and Understanding Workshop (ASRU)}, 2017, pp. 8--15.

\bibitem{dm2019spkbeam2}
M.~Delcroix, K.~Zmolikova, T.~Ochiai, K.~Kinoshita, S.~Araki, and T.~Nakatani, ``Compact network for {S}peaker{B}eam target speaker extraction,'' in \emph{ICASSP 2019 - 2019 IEEE International Conference on Acoustics, Speech and Signal Processing (ICASSP)}, 2019, pp. 6965--6969.

\bibitem{zk2019spkbeam3}
K.~{\v{Z}}mol{\'\i}kov{\'a}, M.~Delcroix, K.~Kinoshita, T.~Ochiai, T.~Nakatani, L.~Burget, and J.~{\v{C}}ernock{\`y}, ``{S}peaker{B}eam: Speaker aware neural network for target speaker extraction in speech mixtures,'' \emph{IEEE Journal of Selected Topics in Signal Processing}, vol.~13, no.~4, pp. 800--814, 2019.

\bibitem{Wang2019VoiceFilterTV}
Q.~Wang, H.~Muckenhirn, K.~Wilson, P.~Sridhar, Z.~Wu, J.~R. Hershey, R.~A. Saurous, R.~J. Weiss, Y.~Jia, and I.~L. Moreno, ``{V}oice{F}ilter: Targeted voice separation by speaker-conditioned spectrogram masking,'' in \emph{Proc. Interspeech 2019}, 2019, pp. 2728--2732.

\bibitem{borsdorf21_interspeech}
M.~Borsdorf, C.~Xu, H.~Li, and T.~Schultz, ``Universal speaker extraction in the presence and absence of target speakers for speech of one and two talkers,'' in \emph{Proc. Interspeech 2021}, 2021, pp. 1469--1473.

\bibitem{Zhang2020XTaSNetRA}
Z.~Zhang, B.~He, and Z.~Zhang, ``{X}-{T}a{SN}et: Robust and accurate time-domain speaker extraction network,'' in \emph{Proc. Interspeech 2020}, 2020, pp. 1421--1425.

\bibitem{roux2019sisnr}
J.~L. Roux, S.~Wisdom, H.~Erdogan, and J.~R. Hershey, ``{SDR} – half-baked or well done?'' in \emph{ICASSP 2019 - 2019 IEEE International Conference on Acoustics, Speech and Signal Processing (ICASSP)}, 2019, pp. 626--630.

\bibitem{shum2013unsupervised}
S.~H. Shum, N.~Dehak, R.~Dehak, and J.~R. Glass, ``Unsupervised methods for speaker diarization: An integrated and iterative approach,'' \emph{IEEE Transactions on Audio, Speech, and Language Processing}, vol.~21, no.~10, pp. 2015--2028, 2013.

\bibitem{shell2014diar}
G.~Sell and D.~Garcia-Romero, ``Speaker diarization with {PLDA} i-vector scoring and unsupervised calibration,'' in \emph{2014 IEEE Spoken Language Technology Workshop (SLT)}, 2014, pp. 413--417.

\bibitem{Mohammed2014sd}
M.~Senoussaoui, P.~Kenny, T.~Stafylakis, and P.~Dumouchel, ``A study of the cosine distance-based mean shift for telephone speech diarization,'' \emph{IEEE/ACM Transactions on Audio, Speech, and Language Processing}, vol.~22, no.~1, pp. 217--227, 2014.

\bibitem{8462628}
Q.~Wang, C.~Downey, L.~Wan, P.~A. Mansfield, and I.~L. Moreno, ``Speaker diarization with {LSTM},'' in \emph{2018 IEEE International Conference on Acoustics, Speech and Signal Processing (ICASSP)}, 2018, pp. 5239--5243.

\bibitem{wang2024overview}
S.~Wang, Z.~Chen, K.~A. Lee, Y.~Qian, and H.~Li, ``Overview of speaker modeling and its applications: From the lens of deep speaker representation learning,'' \emph{IEEE/ACM Transactions on Audio, Speech, and Language Processing}, vol.~32, pp. 4971--4998, 2024.

\bibitem{wang2021scenario}
Y.-X. Wang, J.~Du, M.~He, S.-T. Niu, L.~Sun, and C.-H. Lee, ``Scenario-dependent speaker diarization for {DIHARD-III} challenge,'' in \emph{Proc. Interspeech 2021}, 2021, pp. 3106--3110.

\bibitem{kalda2024pixit}
J.~Kalda, C.~Pagés, R.~Marxer, T.~Alumäe, and H.~Bredin, ``{PixIT}: Joint training of speaker diarization and speech separation from real-world multi-speaker recordings,'' in \emph{The Speaker and Language Recognition Workshop (Odyssey 2024)}, 2024, pp. 115--122.

\bibitem{bando2024neural}
Y.~Bando, T.~Nakamura, and S.~Watanabe, ``Neural blind source separation and diarization for distant speech recognition,'' in \emph{Interspeech 2024}, 2024, pp. 722--726.

\bibitem{wisdom2020unsupervised}
S.~Wisdom, E.~Tzinis, H.~Erdogan, R.~Weiss, K.~Wilson, and J.~Hershey, ``Unsupervised sound separation using mixture invariant training,'' \emph{Advances in neural information processing systems}, vol.~33, pp. 3846--3857, 2020.

\bibitem{kinoshita2021advances}
K.~Kinoshita, M.~Delcroix, and N.~Tawara, ``Advances in integration of end-to-end neural and clustering-based diarization for real conversational speech,'' in \emph{Interspeech 2021}, 2021, pp. 3565--3569.

\bibitem{9414333}
{Kinoshita, Keisuke and Delcroix, Marc and Tawara, Naohiro}, ``Integrating end-to-end neural and clustering-based diarization: Getting the best of both worlds,'' in \emph{ICASSP 2021 - 2021 IEEE International Conference on Acoustics, Speech and Signal Processing (ICASSP)}, 2021, pp. 7198--7202.

\bibitem{9506855}
Y.~Bando, K.~Sekiguchi, Y.~Masuyama, A.~A. Nugraha, M.~Fontaine, and K.~Yoshii, ``Neural full-rank spatial covariance analysis for blind source separation,'' \emph{IEEE Signal Processing Letters}, vol.~28, pp. 1670--1674, 2021.

\bibitem{bando22_interspeech}
Y.~Bando, T.~Aizawa, K.~Itoyama, and K.~Nakadai, ``Weakly-supervised neural full-rank spatial covariance analysis for a front-end system of distant speech recognition,'' in \emph{Interspeech 2022}, 2022, pp. 3824--3828.

\bibitem{graves2006connectionist}
A.~Graves, S.~Fern{\'a}ndez, F.~Gomez, and J.~Schmidhuber, ``Connectionist temporal classification: labelling unsegmented sequence data with recurrent neural networks,'' in \emph{Proceedings of the 23rd international conference on Machine learning}, 2006, pp. 369--376.

\bibitem{he2016deep}
{He, Kaiming and Zhang, Xiangyu and Ren, Shaoqing and Sun, Jian}, ``Deep residual learning for image recognition,'' in \emph{Proceedings of the IEEE conference on computer vision and pattern recognition}, 2016, pp. 770--778.

\bibitem{chung2020defence}
J.~S. Chung, J.~Huh, S.~Mun, M.~Lee, H.-S. Heo, S.~Choe, C.~Ham, S.~Jung, B.-J. Lee, and I.~Han, ``In defence of metric learning for speaker recognition,'' in \emph{Interspeech 2020}, 2020, pp. 2977--2981.

\bibitem{zhou2021resnext}
T.~Zhou, Y.~Zhao, and J.~Wu, ``{ResNeXt and Res2Net} structures for speaker verification,'' in \emph{2021 IEEE Spoken Language Technology Workshop (SLT)}.\hskip 1em plus 0.5em minus 0.4em\relax IEEE, 2021, pp. 301--307.

\bibitem{wang2023wespeaker}
H.~Wang, C.~Liang, S.~Wang, Z.~Chen, B.~Zhang, X.~Xiang, Y.~Deng, and Y.~Qian, ``Wespeaker: A research and production oriented speaker embedding learning toolkit,'' in \emph{ICASSP 2023-2023 IEEE International Conference on Acoustics, Speech and Signal Processing (ICASSP)}.\hskip 1em plus 0.5em minus 0.4em\relax IEEE, 2023, pp. 1--5.

\bibitem{zeinali2019but}
H.~Zeinali, S.~Wang, A.~Silnova, P.~Mat{\v{e}}jka, and O.~Plchot, ``{BUT} system description to {VoxCeleb} speaker recognition challenge 2019,'' \emph{arXiv preprint arXiv:1910.12592}, 2019.

\bibitem{wang24fa_interspeech}
S.~Wang, K.~Zhang, S.~Lin, J.~Li, X.~Wang, M.~Ge, J.~Yu, Y.~Qian, and H.~Li, ``{WeSep}: A scalable and flexible toolkit towards generalizable target speaker extraction,'' in \emph{Interspeech 2024}, 2024, pp. 4273--4277.

\bibitem{panayotov2015librispeech}
V.~Panayotov, G.~Chen, D.~Povey, and S.~Khudanpur, ``{L}ibri{S}peech: An {ASR} corpus based on public domain audio books,'' in \emph{2015 IEEE International Conference on Acoustics, Speech and Signal Processing (ICASSP)}, 2015, pp. 5206--5210.

\bibitem{wichern2019wham}
G.~Wichern, J.~Antognini, M.~Flynn, L.~R. Zhu, E.~McQuinn, D.~Crow, E.~Manilow, and J.~L. Roux, ``{{WHAM}!: Extending Speech Separation to Noisy Environments},'' in \emph{Proc. Interspeech 2019}, 2019, pp. 1368--1372.

\bibitem{li2023an}
\BIBentryALTinterwordspacing
K.~Li, R.~Yang, and X.~Hu, ``An efficient encoder-decoder architecture with top-down attention for speech separation,'' in \emph{The Eleventh International Conference on Learning Representations}, 2023. [Online]. Available: \url{https://openreview.net/forum?id=fzberKYWKsI}
\BIBentrySTDinterwordspacing

\bibitem{yang21c_interspeech}
S.~wen Yang, P.-H. Chi, Y.-S. Chuang, C.-I.~J. Lai, K.~Lakhotia, Y.~Y. Lin, A.~T. Liu, J.~Shi, X.~Chang, G.-T. Lin, T.-H. Huang, W.-C. Tseng, K.~tik Lee, D.-R. Liu, Z.~Huang, S.~Dong, S.-W. Li, S.~Watanabe, A.~Mohamed, and H.~yi~Lee, ``{SUPERB}: Speech processing universal performance benchmark,'' in \emph{Proc. Interspeech 2021}, 2021, pp. 1194--1198.

\bibitem{lspex2022ge}
M.~Ge, C.~Xu, L.~Wang, E.~S. Chng, J.~Dang, and H.~Li, ``{L}-{S}p{E}x: Localized target speaker extraction,'' in \emph{ICASSP 2022 - 2022 IEEE International Conference on Acoustics, Speech and Signal Processing (ICASSP)}, 2022, pp. 7287--7291.

\bibitem{Mark_Martin_2001}
\BIBentryALTinterwordspacing
P.~Mark and A.~Martin, ``2000 {NIST} speaker recognition evaluation,'' 2001. [Online]. Available: \url{https://catalog.ldc.upenn.edu/LDC2001S97}
\BIBentrySTDinterwordspacing

\bibitem{desplanques2020ecapa}
B.~Desplanques, J.~Thienpondt, and K.~Demuynck, ``{ECAPA-TDNN}: Emphasized channel attention, propagation and aggregation in {TDNN} based speaker verification,'' in \emph{Interspeech 2020}, 2020, pp. 3830--3834.

\bibitem{chung2018voxceleb2}
J.~S. Chung, A.~Nagrani, and A.~Zisserman, ``{VoxCeleb2}: Deep speaker recognition,'' in \emph{Interspeech 2018}, 2018, pp. 1086--1090.

\bibitem{stoi}
C.~H. Taal, R.~C. Hendriks, R.~Heusdens, and J.~Jensen, ``A short-time objective intelligibility measure for time-frequency weighted noisy speech,'' in \emph{2010 IEEE International Conference on Acoustics, Speech and Signal Processing}, 2010, pp. 4214--4217.

\bibitem{pesq}
A.~Rix, J.~Beerends, M.~Hollier, and A.~Hekstra, ``Perceptual evaluation of speech quality (pesq)-a new method for speech quality assessment of telephone networks and codecs,'' in \emph{2001 IEEE International Conference on Acoustics, Speech, and Signal Processing. Proceedings (Cat. No.01CH37221)}, vol.~2, 2001, pp. 749--752 vol.2.

\bibitem{wav2vec2020al}
\BIBentryALTinterwordspacing
A.~Baevski, Y.~Zhou, A.~Mohamed, and M.~Auli, ``wav2vec 2.0: A framework for self-supervised learning of speech representations,'' in \emph{Advances in Neural Information Processing Systems}, H.~Larochelle, M.~Ranzato, R.~Hadsell, M.~Balcan, and H.~Lin, Eds., vol.~33.\hskip 1em plus 0.5em minus 0.4em\relax Curran Associates, Inc., 2020, pp. 12\,449--12\,460. [Online]. Available: \url{https://proceedings.neurips.cc/paper_files/paper/2020/file/92d1e1eb1cd6f9fba3227870bb6d7f07-Paper.pdf}
\BIBentrySTDinterwordspacing

\bibitem{hsu2021hubert}
W.-N. Hsu, B.~Bolte, Y.-H.~H. Tsai, K.~Lakhotia, R.~Salakhutdinov, and A.~Mohamed, ``{H}u{BERT}: Self-supervised speech representation learning by masked prediction of hidden units,'' \emph{IEEE/ACM Transactions on Audio, Speech, and Language Processing}, vol.~29, pp. 3451--3460, 2021.

\bibitem{fujita2020neural}
Y.~Fujita, S.~Watanabe, S.~Horiguchi, Y.~Xue, J.~Shi, and K.~Nagamatsu, ``Neural speaker diarization with speaker-wise chain rule,'' \emph{arXiv preprint arXiv:2006.01796}, 2020.

\bibitem{chen23n_interspeech}
Z.~Chen, B.~Han, S.~Wang, and Y.~Qian, ``Attention-based encoder-decoder network for end-to-end neural speaker diarization with target speaker attractor,'' in \emph{Proc. INTERSPEECH 2023}, 2023, pp. 3552--3556.

\end{thebibliography}

\begin{IEEEbiography}[{\includegraphics[width=1in,height=1.25in,clip,keepaspectratio]{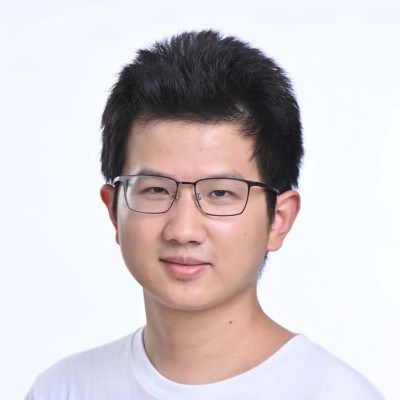}}]{Junyi Ao} (Member, IEEE) received the bachelor’s degree from the Southern University of Science and Technology.
He is a PhD student at the School of Data Science, the Chinese University of Hong Kong, Shenzhen.
He is also the Reviewer of the International Conference on Acoustics, Speech, and Signal Processing, INTERSPEECH, IEEE Transactions on Multimedia, and IEEE Signal Processing Letters.
His research interests include automatic speech recognition, speech pre-training and large language models.
\end{IEEEbiography}

\begin{IEEEbiography}[{\includegraphics[width=1in,height=1.25in,clip,keepaspectratio]{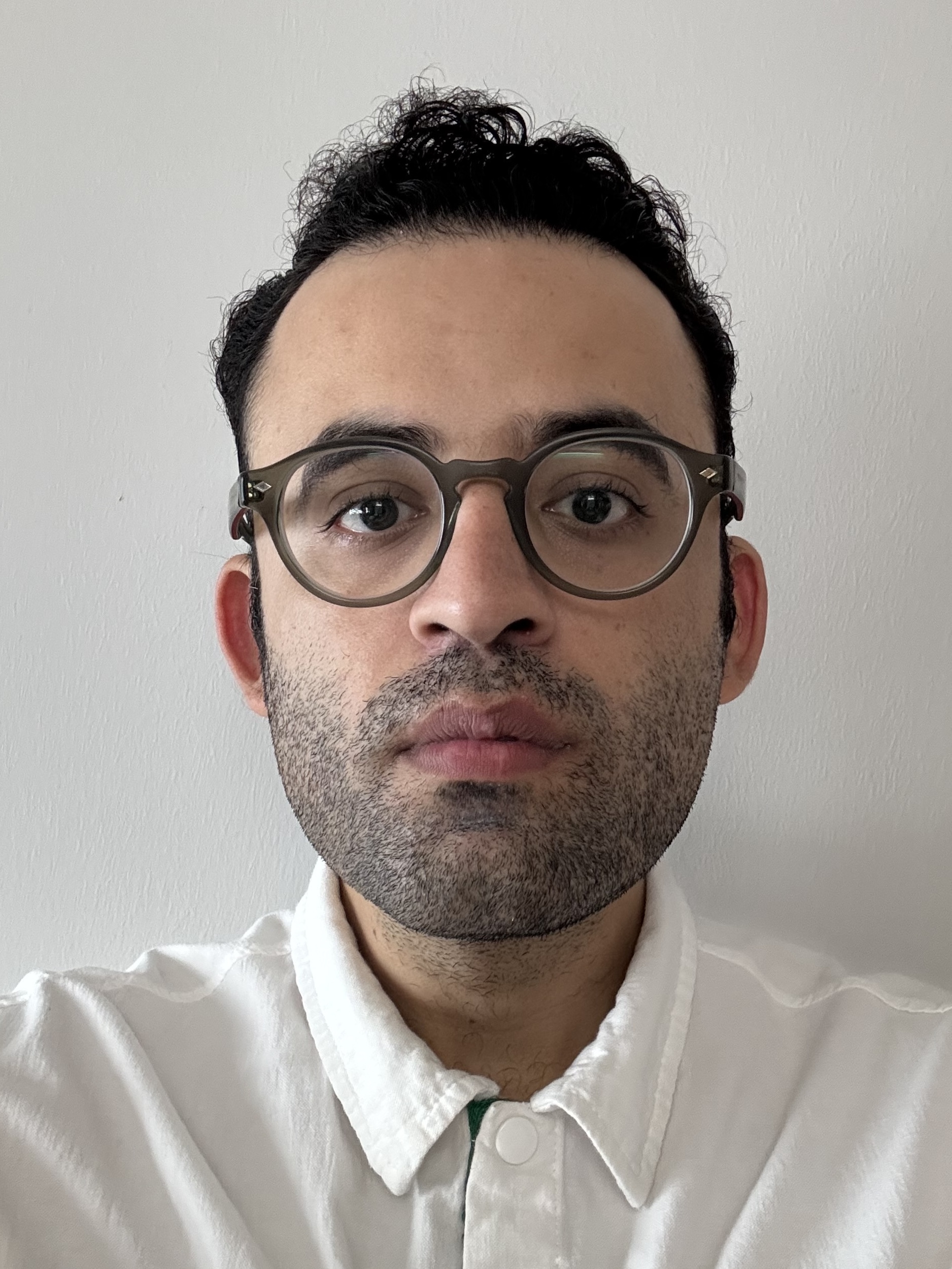}}]{Mehmet Sinan Yıldırım} received the B.Sc. degree from Boğaziçi University, Istanbul, Turkey in 2018. He is currently pursuing the Ph.D. degree in the Electrical and Computer Engineering department at National University of Singapore. His research interests focus on target speaker extraction and target language extraction.
\end{IEEEbiography}

\begin{IEEEbiography}[{\includegraphics[width=1in,height=1.25in,clip,keepaspectratio]{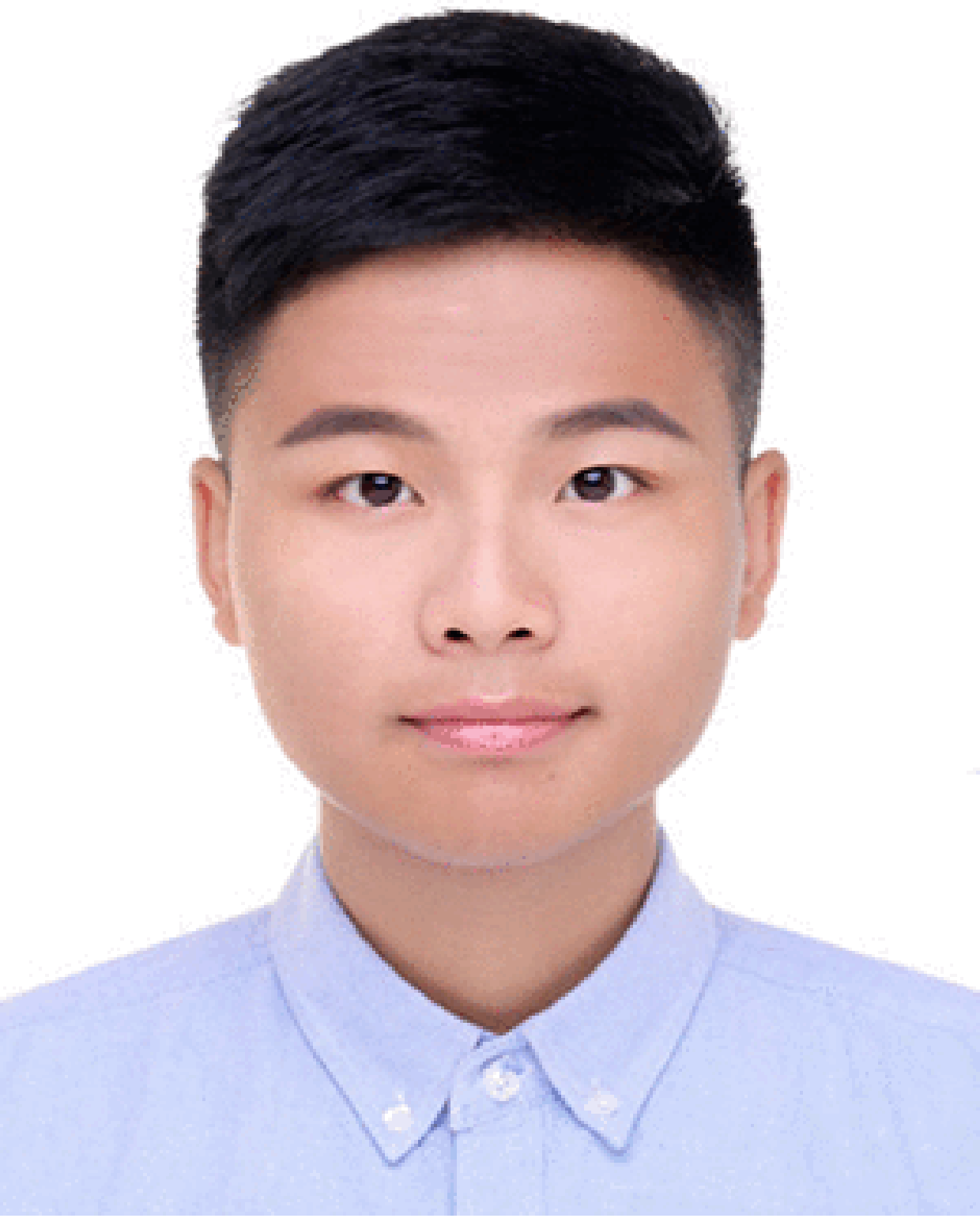}}]{Ruijie Tao} (Member, IEEE) received the B.Eng. degree from Soochow University,
Suzhou, China, in 2018, and the M.Sc. degree in 2019
from the National University of Singapore, Singapore, where he is currently working toward the Ph.D.
degree. He is also the Reviewer of Conference on
Computer Vision and Pattern Recognition, International Conference on Acoustics, Speech, and Signal
Processing, SPL, IEEE OPEN JOURNAL OF SIGNAL
PROCESSING , International Symposium on Chinese
Spoken Language Processing. His research interests
include speaker recognition, speech extraction, and audio-visual representation
learning. 
\end{IEEEbiography}

\begin{IEEEbiography}[{\includegraphics[width=1in,height=1.25in,clip,keepaspectratio]{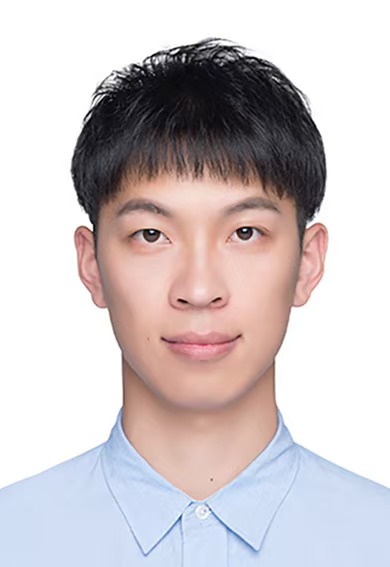}}]{Meng Ge} received the B.Eng. degree in software engineering from Tianjin Polytechnic University, Tianjin, China, in 2015, and received the M.Eng. degree and Ph.D. from the College of Intelligence and Computing of Tianjin University, Tianjin, China,  in 2017 and 2022, respectively. Since 2023, he has been working as a Research Fellow, Saw Swee Hock School of Public Health, National University of Singapore. His research interests include social network analysis, data mining, source separation, and multi-talker robust automatic speech recognition.
\end{IEEEbiography}

\begin{IEEEbiography}[{\includegraphics[width=1in,height=1.25in,clip,keepaspectratio]{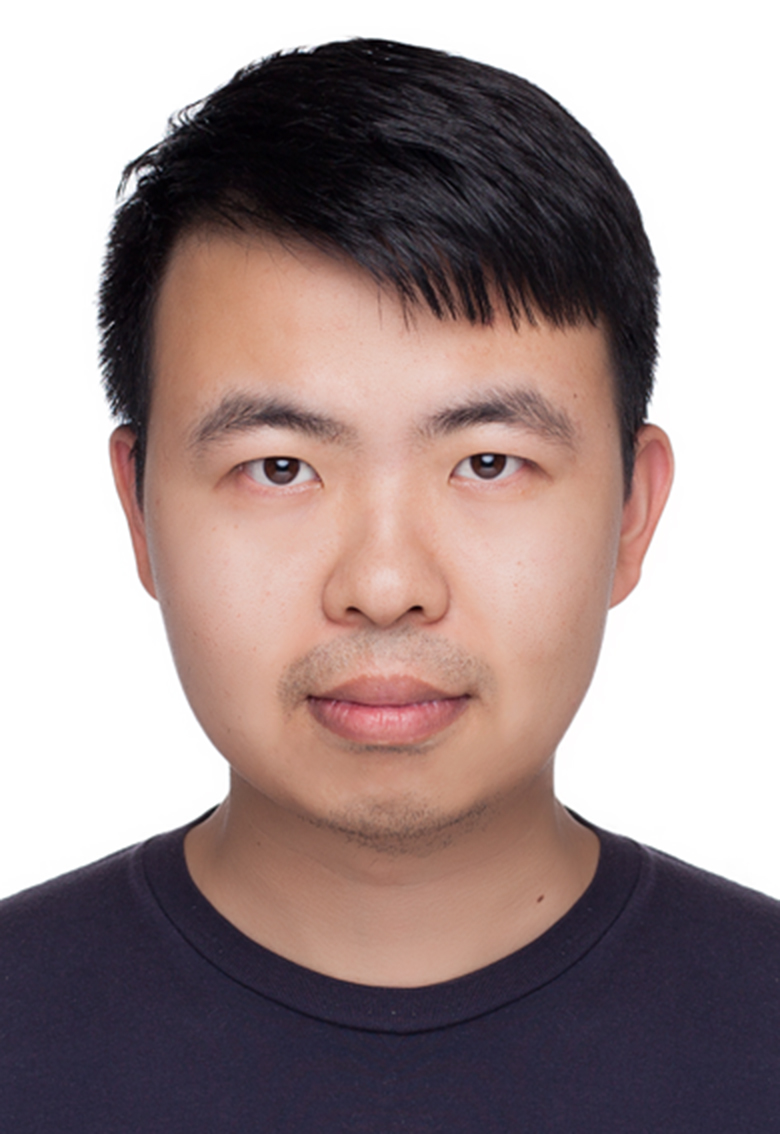}}]{Shuai Wang} (Member, IEEE) obtained his Ph.D. from Shanghai Jiao Tong University (SJTU), in 2020. Currently, he serves as a research scientist at the Shenzhen Research Institute of Big Data (SRIBD). He is also an adjunct assistant professor at School of Data Science, Chinese University of Hong Kong, Shenzhen (CUHK-SZ). Prior to that, Dr. Wang worked at Tencent as a senior application scientist, leading a group working on speaker recognition, voice conversion and speech synthesis. Dr. Wang has published more than 40 papers on the topic of speaker modeling. He was the recipient of IEEE Ganesh N. Ramaswamy Memorial Student Grant Award (ICASSP2018). He was also the main contributor to the winning systems of VoxSRC 2019 and DIHARD 2019.  He is a member of ISCA, SPS and IEEE, searving as a regular reviewer for conferences and journals including ICASSP, INTERSPEECH, ASRU, TASLP and CSL. He initiated the popular ``WeSpeaker" and ``WeSep" projects, utilized by numerous research groups across academia and industry. 
\end{IEEEbiography}

\begin{IEEEbiography}[{\includegraphics[width=1in,height=1.25in,clip,keepaspectratio]{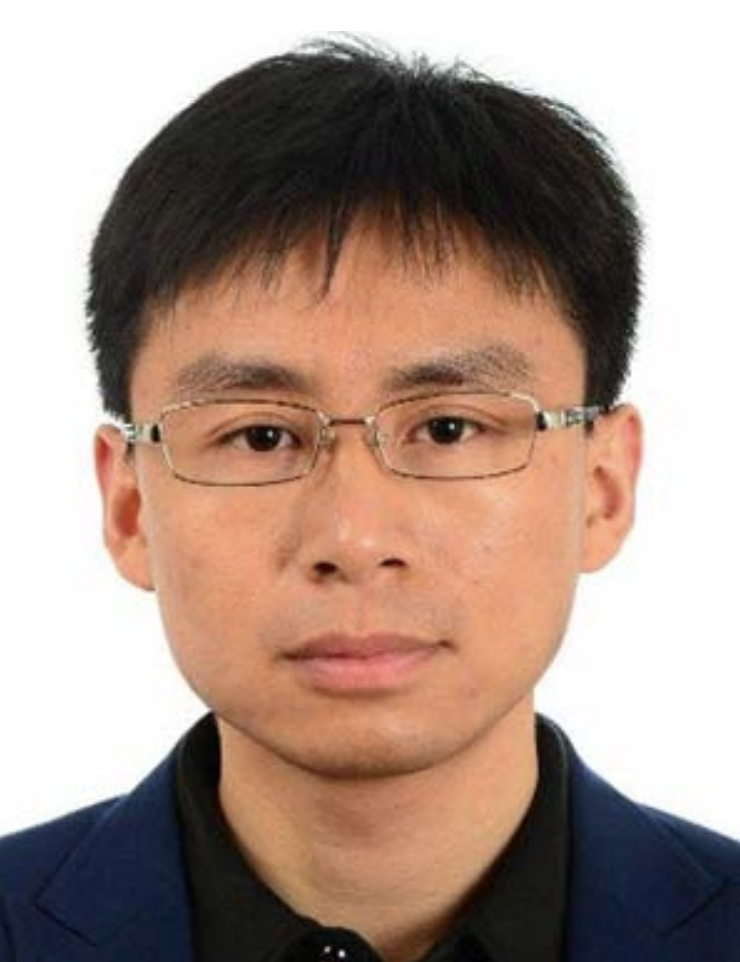}}]{Yanmin Qian}
(Senior Member, IEEE) 
received the B.S. degree from the Department of Electronic and Information Engineering, Huazhong University of Science and Technology, Wuhan, China, in 2007, and the Ph.D. degree from the Department of Electronic Engineering, Tsinghua University, Beijing, China, in 2012. Since 2013, he has been with the Department of Computer Science and Engineering, Shanghai Jiao Tong University, Shanghai, China, where he is currently a Full Professor. From 2015 to 2016, he was an Associate Research with the Speech Group, Cambridge University Engineering Department, Cambridge, U.K. He has authored or coauthored more than 300 papers in peer-reviewed journals and conferences on speech and language processing, including T-ASLP, Speech Communication, ICASSP, INTERSPEECH and ASRU. He has applied for more than 80 Chinese and American patents and won 6 championships of international challenges. His current research interests include speech and audio signal processing, automatic speech recognition and translation, speaker and language recognition, speech separation and enhancement, music generation and understanding, speech emotion perception, multimodal information processing, natural language understanding, deep learning and multi-media signal processing. He was the recipient of several top academic awards in China, including Chang Jiang Scholars Program of the Ministry of Education, Excellent Youth Fund of the National Natural Science Foundation of China, and the First Prize of Wu Wenjun Artificial Intelligence Science and Technology Award (First Completion). He was also the recipient of several awards from international research committee, including the Best Paper Award in Speech Communication and Best Paper Award from IEEE ASRU in 2019. He is also the Member of IEEE Signal Processing Society Speech and Language Technical Committee.
\end{IEEEbiography}

\begin{IEEEbiography}[{\includegraphics[width=1in,height=1.25in,clip,keepaspectratio]{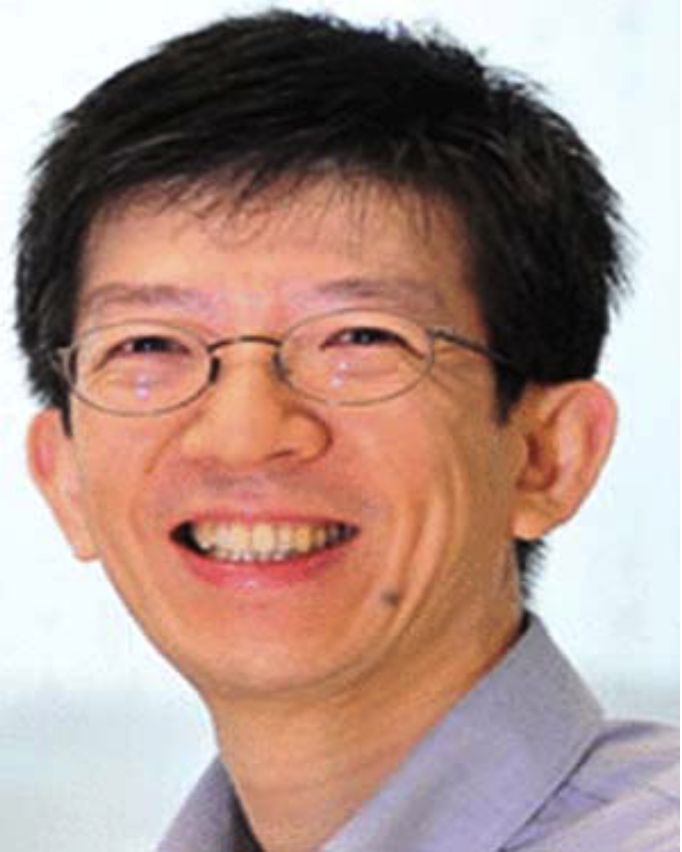}}]{Haizhou Li} (Fellow, IEEE) received the B.Sc.,
M.Sc., and Ph.D. degrees in electrical and electronic
engineering from the South China University of Technology, Guangzhou, China, in 1984, 1987, and 1990
respectively. He is currently a Presidential Chair
Professor and the Executive Dean of the School of
Data Science, The Chinese University of Hong Kong,
Shenzhen, China. He is also an Adjunct Professor
with the Department of Electrical and Computer Engineering, National University of Singapore, Singapore. Prior to that, he taught with The University
of Hong Kong, Hong Kong, during 1988–1990, and South China University
of Technology, during 1990–1994. He was a Visiting Professor with CRIN,
France, during 1994–1995, Research Manager with the AppleISS Research
Centre during 1996–1998, the Research Director with Lernout \& Hauspie
Asia Pacific during 1999–2001, Vice President with InfoTalk Corporation Ltd.
during 2001–2003, and Principal Scientist and Department Head of human
language technology with the Institute for Infocomm Research, Singapore,
during 2003–2016. His research interests include automatic speech recognition,
speaker and language recognition, natural language processing. Dr. Li was
the Editor-in-Chief of IEEE/ACM T RANSACTIONS ON AUDIO , SPEECH AND
LANGUAGE PROCESSING during 2015–2018, an elected Member of IEEE Speech
and Language Processing Technical Committee during 2013–2015, the President
of the International Speech Communication Association during 2015–2017,
President of Asia Pacific Signal and Information Processing Association during
2015–2016, and President of Asian Federation of Natural Language Processing
during 2017–2018. Since 2012, he has been a Member of the Editorial Board
of Computer Speech and Language. He was the General Chair of ACL 2012,
INTERSPEECH 2014, ASRU 2019 and ICASSP 2022. Dr. Li is a Fellow of
the ISCA, and a Fellow of the Academy of Engineering Singapore. He was the
recipient of the National Infocomm Award 2002, and President’s Technology
Award 2013 in Singapore. He was named one of the two Nokia Visiting
Professors in 2009 by the Nokia Foundation, and U Bremen Excellence Chair
Professor in 2019.
\end{IEEEbiography}

\vfill

\end{document}